\pdfoutput=1

\documentclass[12pt,a4paper]{article}

\usepackage{ifthen} 
\newboolean{pdflatex}
\setboolean{pdflatex}{true} 

\newboolean{articletitles}
\setboolean{articletitles}{true} 

\newboolean{uprightparticles}
\setboolean{uprightparticles}{false} 

\newboolean{inbibliography}
\setboolean{inbibliography}{false} 

\usepackage{rotating} 
\usepackage{microtype}
\usepackage{lineno}  
\usepackage{xspace} 
\usepackage{caption} 

\usepackage{graphicx}  
\usepackage{color}
\usepackage{colortbl}
\graphicspath{{./figs/}} 

\usepackage{amsmath} 
\usepackage{amssymb}
\usepackage{amsfonts}
\usepackage{upgreek} 

\newcommand*\patchAmsMathEnvironmentForLineno[1]{%
\expandafter\let\csname old#1\expandafter\endcsname\csname #1\endcsname
\expandafter\let\csname oldend#1\expandafter\endcsname\csname
end#1\endcsname
 \renewenvironment{#1}%
   {\linenomath\csname old#1\endcsname}%
   {\csname oldend#1\endcsname\endlinenomath}%
}
\newcommand*\patchBothAmsMathEnvironmentsForLineno[1]{%
  \patchAmsMathEnvironmentForLineno{#1}%
  \patchAmsMathEnvironmentForLineno{#1*}%
}
\AtBeginDocument{%
\patchBothAmsMathEnvironmentsForLineno{equation}%
\patchBothAmsMathEnvironmentsForLineno{align}%
\patchBothAmsMathEnvironmentsForLineno{flalign}%
\patchBothAmsMathEnvironmentsForLineno{alignat}%
\patchBothAmsMathEnvironmentsForLineno{gather}%
\patchBothAmsMathEnvironmentsForLineno{multline}%
\patchBothAmsMathEnvironmentsForLineno{eqnarray}%
}

\usepackage{hyperref}    
\usepackage[all]{hypcap} 

\def\lhcb {\mbox{LHCb}\xspace}
\def\atlas  {\mbox{ATLAS}\xspace}
\def\cms    {\mbox{CMS}\xspace}

\def\lhc    {\mbox{LHC}\xspace}

\def\MagUp {\mbox{\em Mag\kern -0.05em Up}\xspace}

\ifthenelse{\boolean{uprightparticles}}%
{

 \def\PDelta      {\ensuremath{\Delta}\xspace}                 
 \def\PXi      {\ensuremath{\Xi}\xspace}                 
 \def\PLambda      {\ensuremath{\Lambda}\xspace}                 
 \def\PSigma      {\ensuremath{\Sigma}\xspace}                 
 \def\POmega      {\ensuremath{\Omega}\xspace}                 
 \def\PUpsilon      {\ensuremath{\Upsilon}\xspace}                 
 
 \def\PB      {\ensuremath{\mathrm{B}}\xspace}                 
                  
 \def\PD      {\ensuremath{\mathrm{D}}\xspace}

 \def\PK      {\ensuremath{\mathrm{K}}\xspace}

 \def\PW      {\ensuremath{\mathrm{W}}\xspace}

 \def\PZ      {\ensuremath{\mathrm{Z}}\xspace}                 
                  
 \def\Pb      {\ensuremath{\mathrm{b}}\xspace}                 
 \def\Pc      {\ensuremath{\mathrm{c}}\xspace}

 \def\Pi      {\ensuremath{\mathrm{i}}\xspace}

}
{

 \mathchardef\PDelta="7101
 \mathchardef\PXi="7104
 \mathchardef\PLambda="7103
 \mathchardef\PSigma="7106
 \mathchardef\POmega="710A
 \mathchardef\PUpsilon="7107
                  
 \def\PB      {\ensuremath{B}\xspace}                 
                  
 \def\PD      {\ensuremath{D}\xspace}

 \def\PK      {\ensuremath{K}\xspace}

 \def\PW      {\ensuremath{W}\xspace}

 \def\PZ      {\ensuremath{Z}\xspace}                 
                  
 \def\Pb      {\ensuremath{b}\xspace}                 
 \def\Pc      {\ensuremath{c}\xspace}

 \def\Pi      {\ensuremath{i}\xspace}

}

\makeatletter
\ifcase \@ptsize \relax
  \newcommand{\miniscule}{\@setfontsize\miniscule{4}{5}}
\or
  \newcommand{\miniscule}{\@setfontsize\miniscule{5}{6}}
\or
  \newcommand{\miniscule}{\@setfontsize\miniscule{5}{6}}
\fi
\makeatother

\DeclareRobustCommand{\optbar}[1]{\shortstack{{\miniscule (\rule[.5ex]{1.25em}{.18mm})}
  \\ [-.7ex] $#1$}}

\def\Wp     {{\ensuremath{\PW^+}}\xspace}
\def\Wm     {{\ensuremath{\PW^-}}\xspace}

\def\cquark    {{\ensuremath{\Pc}}\xspace}

\def\bquark    {{\ensuremath{\Pb}}\xspace}

  \def\Kbar    {{\kern 0.2em\overline{\kern -0.2em \PK}{}}\xspace}

\def\KorKbar    {\kern 0.18em\optbar{\kern -0.18em K}{}\xspace}

  \def\Dbar    {{\kern 0.2em\overline{\kern -0.2em \PD}{}}\xspace}

\def\DorDbar    {\kern 0.18em\optbar{\kern -0.18em D}{}\xspace}

\def\Bbar    {{\ensuremath{\kern 0.18em\overline{\kern -0.18em \PB}{}}}\xspace}

\def\BorBbar    {\kern 0.18em\optbar{\kern -0.18em B}{}\xspace}

  \def\Y#1S{\ensuremath{\PUpsilon{(#1S)}}\xspace}

\def\Lbar        {{\ensuremath{\kern 0.1em\overline{\kern -0.1em\PLambda}}}\xspace}
\def\LorLbar    {\kern 0.18em\optbar{\kern -0.18em \PLambda}{}\xspace}

\def\AT#1     {\ensuremath{A_{\mathrm{T}}^{#1}}\xspace}           

\def\C#1      {\ensuremath{\mathcal{C}_{#1}}\xspace}                       
\def\Cp#1     {\ensuremath{\mathcal{C}_{#1}^{'}}\xspace}                    
\def\Ceff#1   {\ensuremath{\mathcal{C}_{#1}^{\mathrm{(eff)}}}\xspace}        
\def\Cpeff#1  {\ensuremath{\mathcal{C}_{#1}^{'\mathrm{(eff)}}}\xspace}       
\def\Ope#1    {\ensuremath{\mathcal{O}_{#1}}\xspace}                       
\def\Opep#1   {\ensuremath{\mathcal{O}_{#1}^{'}}\xspace}                    

\newcommand{\tev}{\ifthenelse{\boolean{inbibliography}}{\ensuremath{~T\kern -0.05em eV}\xspace}{\ensuremath{\mathrm{\,Te\kern -0.1em V}}}\xspace}
\newcommand{\gev}{\ensuremath{\mathrm{\,Ge\kern -0.1em V}}\xspace}
\newcommand{\mev}{\ensuremath{\mathrm{\,Me\kern -0.1em V}}\xspace}
\newcommand{\kev}{\ensuremath{\mathrm{\,ke\kern -0.1em V}}\xspace}
\newcommand{\ev}{\ensuremath{\mathrm{\,e\kern -0.1em V}}\xspace}
\newcommand{\gevc}{\ensuremath{{\mathrm{\,Ge\kern -0.1em V\!/}c}}\xspace}
\newcommand{\mevc}{\ensuremath{{\mathrm{\,Me\kern -0.1em V\!/}c}}\xspace}
\newcommand{\gevcc}{\ensuremath{{\mathrm{\,Ge\kern -0.1em V\!/}c^2}}\xspace}
\newcommand{\gevgevcccc}{\ensuremath{{\mathrm{\,Ge\kern -0.1em V^2\!/}c^4}}\xspace}
\newcommand{\mevcc}{\ensuremath{{\mathrm{\,Me\kern -0.1em V\!/}c^2}}\xspace}

\def\mum  {\ensuremath{{\,\upmu\rm m}}\xspace}

\def\gsim{{~\raise.15em\hbox{$>$}\kern-.85em
          \lower.35em\hbox{$\sim$}~}\xspace}
\def\lsim{{~\raise.15em\hbox{$<$}\kern-.85em
          \lower.35em\hbox{$\sim$}~}\xspace}

\def\ptot       {\mbox{$p$}\xspace}
\def\pt         {\mbox{$p_{\rm T}$}\xspace}

\def\fewz       {\mbox{\textsc{Fewz}}\xspace}
\def\dynnlo       {\mbox{\textsc{DYNNLO}}\xspace}

\def\geant      {\mbox{\textsc{Geant4}}\xspace}

\def\herwig     {\mbox{\textsc{Herwig}}\xspace}
\def\herwiri     {\mbox{\textsc{Herwiri}}\xspace}
\def\mc@nlo     {\mbox{\textsc{MC@NLO}}\xspace}

\def\powheg     {\mbox{\textsc{Powheg}}\xspace}
\def\pythia     {\mbox{\textsc{Pythia}}\xspace}
\def\resbos     {\mbox{\textsc{ResBos}}\xspace}

\def\tell1  {TELL1\xspace}
\def\ukl1   {UKL1\xspace}

\usepackage{multirow}
\usepackage{bm}
\usepackage{cite} 
\usepackage{mciteplus}

\begin{document}

\renewcommand{\thefootnote}{\fnsymbol{footnote}}
\setcounter{footnote}{1}


\begin{titlepage}
\pagenumbering{roman}

\vspace*{-1.5cm}
\centerline{\large EUROPEAN ORGANIZATION FOR NUCLEAR RESEARCH (CERN)}
\vspace*{1.5cm}
\hspace*{-0.5cm}
\begin{tabular*}{\linewidth}{lc@{\extracolsep{\fill}}r}
\ifthenelse{\boolean{pdflatex}}
{\vspace*{-1.5cm}\mbox{\!\!\!\includegraphics[width=.14\textwidth]{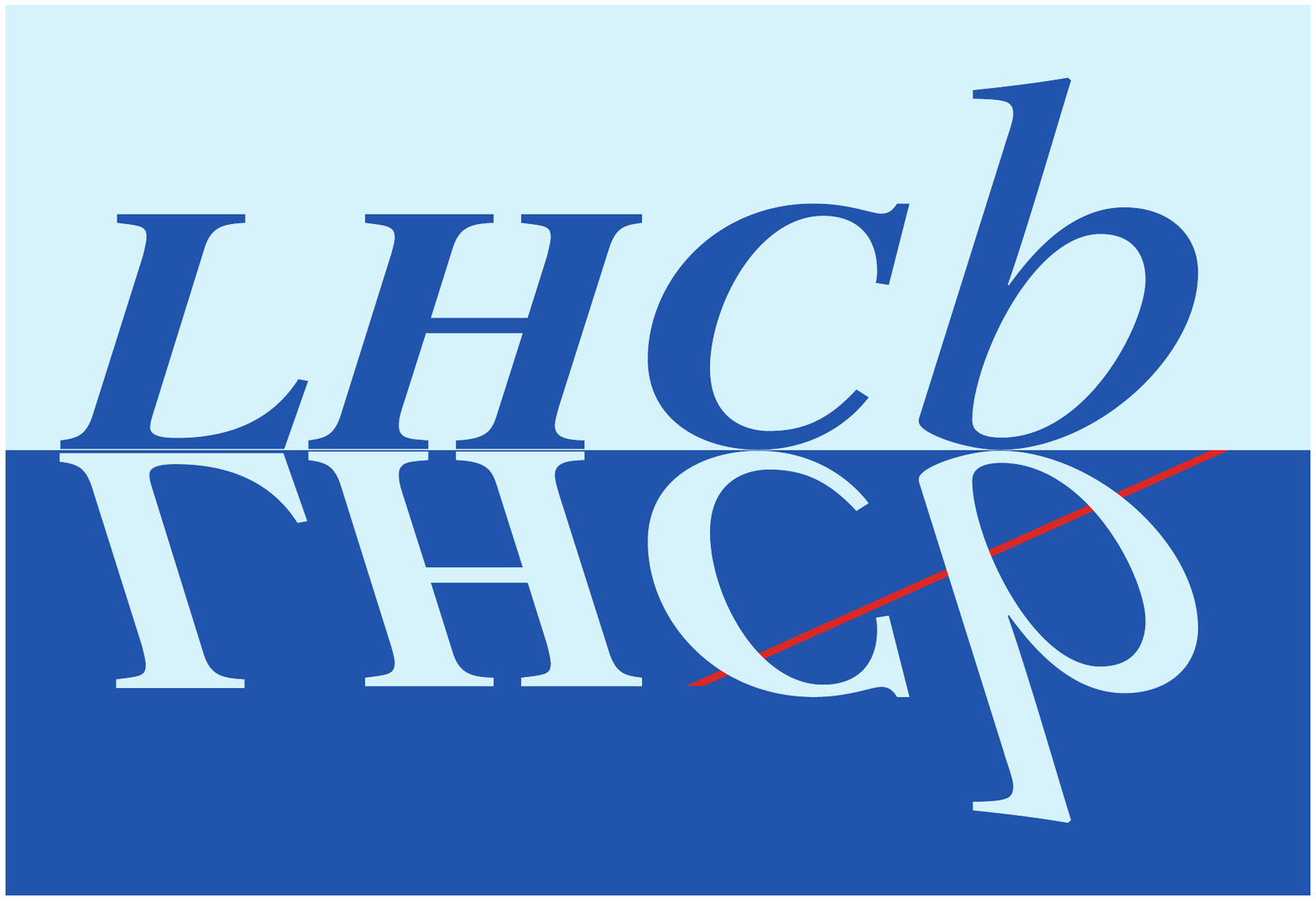}} & &}%
{\vspace*{-1.5cm}\mbox{\!\!\!\includegraphics[width=.12\textwidth]{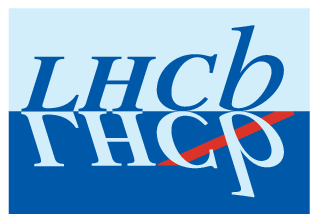}} & &}%
\\
 & & CERN-PH-EP-2015-102 \\  
 & & LHCb-PAPER-2015-001 \\  
 & & Aug 20, 2015 \\ 
 & & \\
\end{tabular*}

\vspace*{0.85cm}

{\bf\boldmath\huge
\begin{center}
Measurement of the forward \PZ boson \\ production cross-section \\ in $pp$ collisions at $\sqrt{s}$ = 7 \tev
\end{center}
}

\vspace*{0.15cm}

\begin{center}
The LHCb collaboration\footnote{Authors are listed at the end of this paper.}
\end{center}

\vspace{0.25cm}
\begin{abstract}
  \noindent
A measurement of the production cross-section for \PZ bosons that decay to muons is presented. 
The data were recorded by the \lhcb detector during $pp$ collisions at a centre-of-mass 
energy of 7 \tev, and correspond to an integrated luminosity of 1.0 fb$^{-1}$.
The cross-section is 
measured for muons in the 
pseudorapidity range $2.0 \nolinebreak<\nolinebreak \eta \nolinebreak<\nolinebreak 4.5$ with 
transverse momenta $\pt \nolinebreak>\nolinebreak 20\gevc$. The dimuon mass  
is restricted to  
$60 \nolinebreak<\nolinebreak M_{\mu^{+}\mu^{-}} \nolinebreak<\nolinebreak 120\gevcc$. The measured cross-section is
$$\sigma_{Z\rightarrow\mu^{+}\mu^{-}} = (76.0 \pm 0.3 \pm 0.5 \pm 1.0 \pm 1.3) \, \text{pb}$$
where the uncertainties are due to the sample size, systematic effects, the beam energy and the luminosity.\enspace
This result is in good agreement with 
theoretical predictions at next-to-next-to-leading order in perturbative quantum
chromodynamics. The cross-section is also measured differentially as a function of kinematic variables 
of the \PZ boson. 
Ratios of the production cross-sections of electroweak bosons are presented
using updated \lhcb measurements of \PW boson production. 
A precise test of the Standard Model is provided by the measurement of the ratio
$$\frac{\sigma_{W^{+}\rightarrow\mu^{+}\nu_{\mu}} + \sigma_{W^{-}\rightarrow\mu^{-}\bar{\nu}_{\mu}}}{\sigma_{\PZ\rightarrow\mu^{+}\mu^{-}}} = 20.63\pm0.09\pm0.12\pm0.05,$$ 
where the uncertainty due to luminosity cancels.
  
\end{abstract}

\vspace*{0.3cm}

\begin{center}
  Published in JHEP \ 08 \ (2015) \ 039 
\end{center}

\vspace{\fill}

{\footnotesize 
\centerline{\copyright~CERN on behalf of the \lhcb collaboration, licence \href{http://creativecommons.org/licenses/by/4.0/}{CC-BY-4.0}.}}

\end{titlepage}


\newpage
\setcounter{page}{2}
\mbox{~}
%
%
%
%

\cleardoublepage


\renewcommand{\thefootnote}{\arabic{footnote}}
\setcounter{footnote}{0}



\pagestyle{plain} 
\setcounter{page}{1}
\pagenumbering{arabic}


%


\section{Introduction}
\label{sec:Introduction}

Measurements of the total and differential cross-sections for the production of \PZ 
bosons
in $pp$ collisions test the Standard Model (SM) and 
provide constraints on parton density functions (PDFs) of the proton.\footnote{Throughout this article \PZ represents both 
resonant production of \PZ bosons and off-mass-shell photons.}
Theoretical predictions for these cross-sections are available at 
next-to-next-to-leading order (NNLO) 
in perturbative quantum chromodynamics (pQCD)~\cite{nnlopQCD1,nnlopQCD2,nnlopQCD3,nnlopQCD4,nnlopQCD5}. The dominant uncertainty on these predictions 
reflects the uncertainties on the PDFs, 
which vary as functions of the kinematic variables studied. 
The forward acceptance of the \lhcb detector allows the PDFs to be 
constrained at Bjorken-$x$ values down to 10$^{-4}$~\cite{LHCbPDF}.
Ratios of the \PW and \PZ cross-sections provide precise tests of the SM as the sensitivity to the PDFs in the theoretical calculations is reduced 
and many of the experimental uncertainties cancel.

\lhcb has measured the \PZ boson production cross-section at $\sqrt{s}$= 7 \tev using decays to muon pairs in a data set 
corresponding to
37 pb$^{-1}$~\cite{lhcbwz}, and using electron~\cite{ze} and tau lepton~\cite{ztau} pairs in a data set of 1.0 fb$^{-1}$. 
Production cross-sections of \PW bosons and the $W^{+}$/$W^{-}$ cross-section ratio
have been measured in the muon channel~\cite{wmu} with the  $1.0\nolinebreak\text{ fb}^{-1}$ data set. Similar measurements 
have also been performed by the \atlas~\cite{atlaswz} 
and \cms~\cite{cmswz} collaborations.

The analysis described here is an update of the one described in Ref.~\cite{lhcbwz}, 
using a total integrated luminosity of about 1.0 fb$^{-1}$. 
This increases statistical precision and allows better control of systematic 
uncertainties, with the result that the 
total uncertainties on the measurements are 
significantly reduced.
Measurements are performed for muons with transverse momentum 
$\pt > 20\gevc$ and pseudorapidity in the range $2.0 < \eta < 4.5$. In the case of 
\PZ boson measurements, the invariant 
mass of the two muons is required to be in the range $60 < M_{\mu^{+}\mu^{-}} < 120\gevcc$. 
These kinematic requirements define the fiducial region of the measurement and 
in this article are referred to as the fiducial requirements. 
Total cross-sections are presented as well as differential 
cross-sections as functions of the \PZ boson 
rapidity $y_{Z}$, $p_{T,Z}$ and $\phi^{*}_{Z}$. Here 
$\phi^{*}_{Z}$ is defined as~\cite{phistar}
\begin{equation}
 \phi^{*}_{Z} \equiv \frac{\tan{(\phi_{\rm acop}/2)}}{\cosh{(\Delta\eta}/2)}.
\end{equation}
The angle $\phi_{\rm acop} = \pi - |\Delta\phi|$ depends on the  
difference $\Delta\phi$ in azimuthal angle between the two muon momenta, 
while the difference between their pseudorapidities is denoted by $\Delta\eta$.

The \PW boson cross-sections given in Ref.~\cite{wmu} are re-evaluated using a more precise 
determination of the event trigger efficiency. The cross-sections are presented as a function of the $\eta$ of 
the muon from the \PW boson decay. The values presented here supersede those of Ref.~\cite{wmu}. 

This paper is organised as follows:
Section~\ref{sec:Detector} describes the \lhcb detector;
Sections~\ref{sec:selection} and~\ref{sec:cs} detail the selection of \PZ boson candidates,
the \PZ boson cross-section definition and relevant sources of systematic uncertainty;
Section~\ref{sec:results} presents the results
and Section~\ref{sec:conclusions} concludes the paper.
Appendices~\ref{sec:difftab} and~\ref{sec:correlation} provide tables of differential cross-sections and 
correlations between these measurements.

\section{Detector and data set}
\label{sec:Detector}

The \lhcb detector~\cite{Alves:2008zz,LHCb-DP-2014-002} is a single-arm forward
spectrometer covering the \mbox{pseudorapidity} range $2<\eta <5$,
designed for the study of particles containing \bquark or \cquark
quarks. The detector includes a high-precision tracking system
consisting of a silicon-strip vertex detector surrounding the $pp$
interaction region~\cite{LHCb-DP-2014-001}, a large-area silicon-strip detector located
upstream of a dipole magnet with a bending power of about
$4{\rm\,Tm}$, and three stations of silicon-strip detectors and straw
drift tubes~\cite{LHCb-DP-2013-003} placed downstream of the magnet.
The tracking system provides a measurement of momentum, \ptot, of charged particles with
a relative uncertainty that varies from 0.5\% at low momentum to 1.0\% at 200\gevc.
The minimum distance of a track to a primary vertex, the impact parameter, is measured with a resolution of $(15+29/\pt)\mum$,
where \pt is the component of the momentum transverse to the beam, in \gevc.
Different types of charged hadrons are distinguished using information
from two ring-imaging Cherenkov detectors~\cite{LHCb-DP-2012-003}.
Photons, electrons and hadrons are identified by a calorimeter system consisting of a
scintillating-pad detector (SPD), preshower detectors, an electromagnetic
calorimeter and a hadronic calorimeter. Muons are identified by a
system composed of alternating layers of iron and multiwire
proportional chambers~\cite{LHCb-DP-2012-002}.
The online event selection is performed by a trigger~\cite{LHCb-DP-2012-004},
which consists of a hardware stage, based on information from the calorimeter and muon
systems, followed by a software stage, which applies a full event 
reconstruction. A requirement that prevents events with high
occupancy from dominating the processing time of the software trigger is 
also applied.
This is referred to as the global event cut (GEC) in this article.

The measurement presented here is based on $pp$ collision data collected at a centre-of-mass energy of 7 \tev. 
The integrated luminosity amounts to 975 $\pm$ 17 pb$^{-1}$.
The absolute luminosity scale was measured during dedicated data taking periods,
using both Van der Meer scans~\cite{vandermeer} 
and beam-gas imaging methods~\cite{beamgas}. Both
methods give similar results, which are combined to give the final luminosity estimate with an uncertainty of 
$1.7\%$~\cite{lumi2}. 
This analysis uses the same data set as in Ref.~\cite{wmu}.

Several samples of simulated data are produced to estimate contributions from background processes, to cross-check efficiencies and to unfold data
for detector-related effects. The \pythia generator~\cite{pythia,pythia8}, configured as in Ref.~\cite{gauss}
with the CTEQ6L1~\cite{cteq,cteq1} parameterisation for the PDFs, is used to simulate $b\bar{b}$, $c\bar{c}$, $WW$, $t\bar{t}$ 
and \PZ production. All generated events are passed
through a detector simulation based on \geant~\cite{geant}, followed by 
\lhcb-specific trigger emulation and event 
reconstruction.
 
The results of the analysis are compared to theoretical predictions calculated with the \fewz~\cite{fewz,fewzold} generator at NNLO for the 
PDF sets ABM12~\cite{abm12},
CT10~\cite{ct10}, HERA1.5~\cite{h1zeus}, JR09~\cite{jr09}, MSTW08~\cite{mstw08}, and NNPDF3.0~\cite{PDF-NNPDF30}. 
Comparisons are also made to the \resbos~\cite{resbos1,resbos2,resbos3} and \powheg~\cite{powheg} generators 
configured with the CT10 PDF set. \resbos includes 
an approximate NNLO calculation, plus a next-to-next-to-leading logarithm approximation for the resummation of the soft gluon radiation.
\powheg provides a next-to-leading order (NLO) calculation interfaced to a parton shower, in this case performed by \herwig~\cite{GEN-HERWIG1,GEN-HERWIG2}.
The results are also compared to the predictions from \mc@nlo ~\cite{mcnlo1,mcnlo2}, which is interfaced 
with different generators to simulate 
the parton shower. Parton showering is performed using 
\herwig~\cite{GEN-HERWIG1,GEN-HERWIG2}, with different values for the root 
mean-square-deviation  of the intrinsic $k_{T}$, 
and \herwiri~\cite{herwiri1,herwiri2,herwiri3}, which is based on infrared-improved~\cite{IRimp1} DGLAP-CS~\cite{dglap1,dglap2,dglap3,dglap4,cs1,cs2} 
theory. All calculations are performed with the renormalisation and factorisation scales set to the electroweak boson mass. 
Scale uncertainties are estimated by varying these scales by factors of two around the boson 
mass~\cite{scaleUnc}.
Total uncertainties 
correspond to the PDF and $\alpha_{s}$ uncertainties at 68.3$\%$ confidence 
level and scale uncertainties, added in quadrature.

\section{Event selection}
\label{sec:selection}
\begin{figure}[h]
  \begin{center}
    \includegraphics[width=0.75\linewidth]{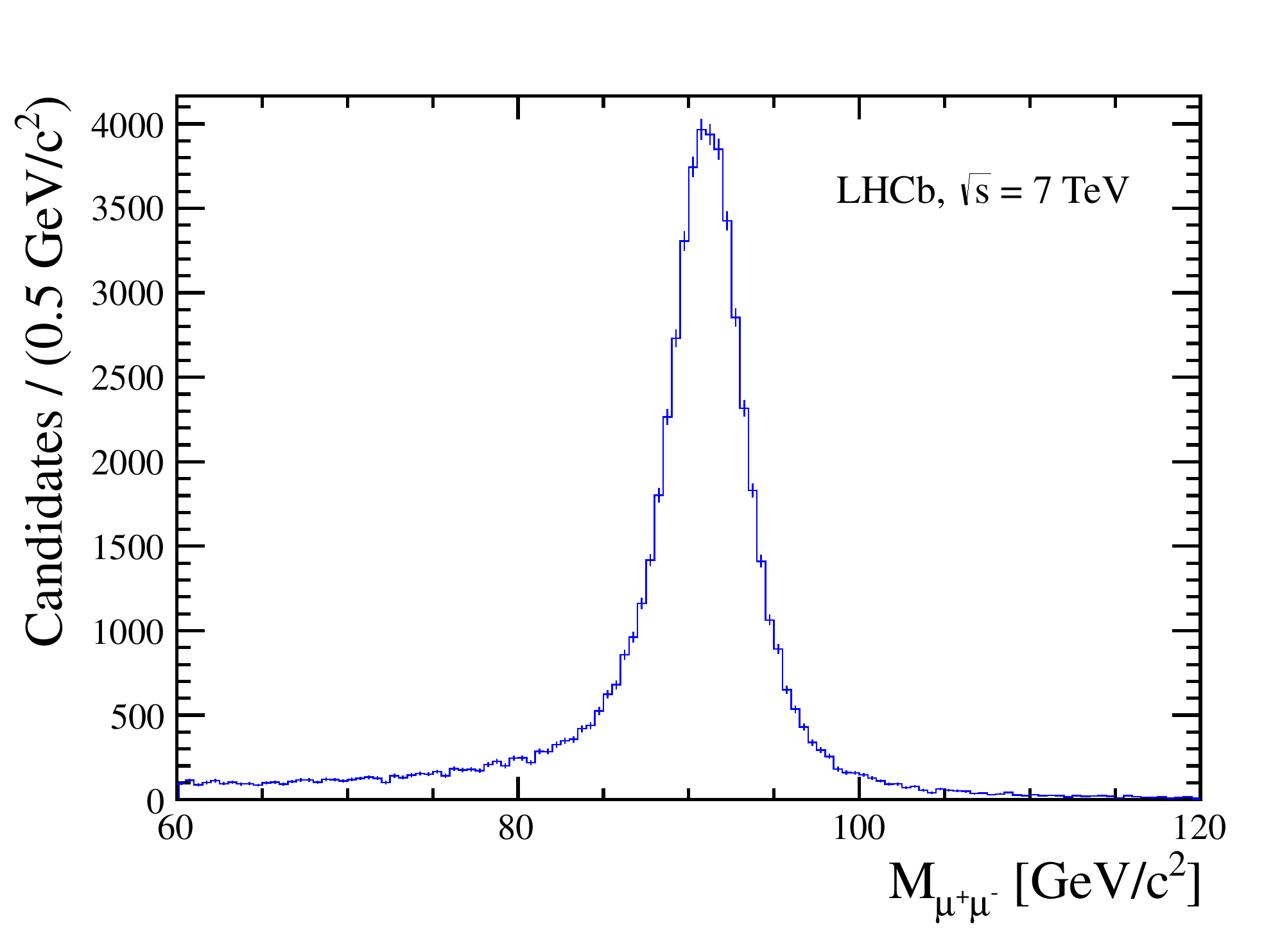}
  \end{center}
  \caption{Invariant mass of dimuon candidates.}
  \label{fig:zmass}
\end{figure}

Events considered in this analysis are selected by the muon trigger.
At the hardware stage, this trigger requires a muon with $\pt > 1.5\gevc$ and imposes an upper limit of 600 hits in the 
SPD sub-detector.
At the software stage, a muon with $\pt > 10\gevc$ is required. The muon track must also satisfy additional track quality 
criteria.
Candidate events are selected by requiring a pair of well-reconstructed particles of opposite
charge, identified as muons, that also pass the fiducial requirements~\cite{lhcbwz}.
In total, 58\,466 \PZ boson candidates are selected and their invariant mass distribution 
is shown in Fig.~\ref{fig:zmass}. 

The background contamination in the candidate sample is low. Five background sources are investigated: 
decays of heavy flavour hadrons, hadron misidentification, $\PZ\nolinebreak\rightarrow\nolinebreak\tau^{+}\tau^{-} $ decays, $t\bar{t}$ and 
$W^{+}W^{-}$ production.
Unlike muons from signal, muons arising from decays of heavy flavour hadrons are neither
directly produced at the primary interaction vertex, nor
are they isolated particles. Using the techniques
from Ref.~\cite{lhcbwz}, the contribution from this background is estimated 
from the data to be $227\pm32$ events, which amounts to 0.4$\%$ of the candidate sample.
The contribution from hadrons that decay in flight or have sufficient energy to traverse the calorimeters 
and be detected in the muon stations is studied 
in randomly triggered data, as described in Ref.~\cite{lhcbwz}, and determined to be $116\pm45$ events, which is 0.2$\%$ of the candidate sample.
Other electroweak and QCD backgrounds are estimated using \pythia 
simulation~\cite{pythia8} and normalised to the measured total 
cross-sections
for $\PZ\rightarrow\tau^{+}\tau^{-}$ decays~\cite{ATLAStautaumeas,CMStautaumeas}, $t\bar{t}$~\cite{ttbarATLAS,ttbarCMS} and 
$W^{+}W^{-}$~\cite{ATLASWWmeas,CMSWWmeas} production. The estimate for these sources is 
$66\pm6$ events, or 0.1$\%$ of the candidate sample. In total, the background is estimated to be $409\pm56$ events, or 0.7$\%$ 
of the candidate sample.

The purity, defined to be the ratio of signal to total candidate events, is 
$\rho$=0.993$\pm$0.002. 
It is assumed to be constant as a function of $y_{Z}$, $p_{T,Z}$ and $\phi^{*}_{Z}$.
A systematic uncertainty, discussed later, is assigned to allow for possible inaccuracies in this assumption.

\section{Cross-section measurement}
\label{sec:cs}

Cross-sections are quoted in the kinematic range defined by the measurement and are corrected 
for quantum electrodynamic (QED) final-state radiation (FSR) in order to 
provide a consistent 
comparison with NLO and NNLO QCD predictions.
No corrections are
applied for initial-state radiation, electroweak effects, nor their 
interplay
with QED effects.
The cross-section in a given bin $i$ of
$y_{Z}$, $p_{T,Z}$ or $\phi^{*}_{Z}$, with both final-state muons
inside the fiducial region, is measured as
\begin{equation}
 \label{eq:cs1}
 \sigma_{\PZ\rightarrow\mu^{+}\mu^{-}}(i) = \frac{\rho}{\mathcal{L}} \ \frac{f_{\rm FSR}(i)}{\varepsilon_{\rm GEC}(i)} 
 \sum_j U_{ij} \left( \sum_k \frac{1}{\varepsilon(\eta_{k}^{\mu^{+}},\eta_{k}^{\mu^{-}})}\right)_{j}.
\end{equation}
The indices $i$ and $j$ run over the bins of the variable under study.  
The index $k$ runs over the candidates contributing to bin $j$.
The total muon reconstruction efficiency for an event is given by
$\varepsilon(\eta_{k}^{\mu^{+}},\eta_{k}^{\mu^{-}})$, which is dependent on the pseudorapidity of the two muons. 
The matrix $U$ corrects the data for bin migrations due to detector resolution effects.
It is determined using an unfolding procedure, which is described in Section~\ref{sec:unfld}.
The efficiency of the
requirement on the number of SPD hits in the hardware trigger is denoted by $\varepsilon_{\rm GEC}$.
The correction factors for QED final-state radiation are denoted 
by $f_{\rm FSR}(i)$ and are
determined for each bin. The integrated luminosity is denoted by $\mathcal{L}$.
Though not entering the expression for the cross-section, an uncertainty due to the beam energy
is assigned to all cross-sections.
More detail on these individual components is given below.
Once the binned cross-sections are determined, they are summed to give the total cross-section
\begin{equation}
 \sigma_{\PZ\rightarrow\mu^{+}\mu^{-}}=\sum_i \sigma_{\PZ\rightarrow\mu^{+}\mu^{-}}(i).
\end{equation}
The most precise estimate of the total cross-section is obtained by summing 
the cross-sections as a function
of rapidity, where uncertainties due to data unfolding are negligible.

\subsection{Muon reconstruction efficiencies}

The data are corrected for efficiency losses due to track reconstruction, muon identification,
and trigger requirements. All efficiencies are determined from data using the techniques detailed in Refs.~\cite{lhcbwz,ze}, 
where the track reconstruction, muon identification, and muon trigger efficiencies
are obtained using tag-and-probe methods on the $\PZ$ resonance. 
The tag and probe tracks are required to satisfy the fiducial requirements.
The tag must be identified as a muon and be consistent with triggering the event, while the
probe is defined so that it is unbiased by the requirement for which the 
efficiency is being measured.
The efficiency is studied as a function of several variables, which 
describe both the muon kinematics
and the detector occupancy. In this analysis the efficiency as a function of muon $\eta$ is used. 
The efficiency in each bin of $\eta$ is defined as the fraction of tag-and-probe
candidate events where the probe satisfies a track reconstruction, identification or trigger requirement. 

The tracking efficiency is determined using probe tracks that are 
reconstructed by combining hits from the muon
stations and the large-area silicon-strip detector. 
The efficiency depends on $\eta$ and varies between 89.5$\%$ and 98.5$\%$ with uncertainties between 
0.4$\%$ and 1.9$\%$.

The muon identification efficiency is determined using probe tracks that are reconstructed without using the muon system.
The efficiency depends on $\eta$ and varies between 91.3$\%$ and 99.2$\%$ with
uncertainties between 0.1$\%$ and 0.9$\%$.        

The single-muon trigger efficiency is determined using reconstructed muons 
as probes. 
The efficiency depends on $\eta$ and varies between 71.6$\%$ and 82.0$\%$ with
uncertainties between 0.5$\%$ and 1.2$\%$.
Since only one muon candidate is required for the event to pass the trigger requirements, 
the overall trigger efficiency for the analysis is about 95$\%$. 

The efficiency to reconstruct any given event is taken to be the product of the three individual
efficiencies and determined on an event-by-event basis as a function of muon $\eta$,
\begin{equation}
 \varepsilon(\mu^{+},\mu^{-}) = \varepsilon_{\rm trk}^{\mu^{+}} \cdot \varepsilon_{\rm trk}^{\mu^{-}} \cdot \varepsilon_{\rm id}^{\mu^{+}} \cdot \varepsilon_{\rm id}^{\mu^{-}} \cdot \left( \varepsilon_{\rm trg}^{\mu^{+}} + \varepsilon_{\rm trg}^{\mu^{-}} - \varepsilon_{\rm trg}^{\mu^{+}} \cdot \varepsilon_{\rm trg}^{\mu^{-}} \right). 
\label{eq:effeqn}
\end{equation}
In Equation~\ref{eq:effeqn}, the efficiency $\varepsilon$ is written explicitly in terms of the muon tracking ($\varepsilon_{\rm trk}$), identification ($\varepsilon_{\rm id}$) and
trigger ($\varepsilon_{\rm trg}$) efficiencies. 
The average reconstruction efficiency for the analysis is about 85$\%$.
Effects that correlate the efficiency of the two muons are considered, but these are negligible at the current level of 
precision. 

\subsection{GEC efficiency}
\label{sec:gec}
The GEC efficiency is the efficiency of the SPD multiplicity limit at 600 hits in the muon trigger.
This efficiency is evaluated from data using two independent methods. The first
exploits the fact that the SPD multiplicities of single $pp$ interactions are always below the 600 hit threshold.
The expected SPD multiplicity distribution of signal events is constructed by adding the multiplicities
of signal events in single $pp$ interactions to the multiplicities of randomly triggered events, as in Ref.~\cite{lhcbwz}.
The convolution of the distributions extends to values above 600 hits, 
and the fraction of events that the trigger rejects can be determined.
The second method consists of fitting
the SPD multiplicity distribution and extrapolating the fit function to determine the fraction of events that are rejected,
as in Ref.~\cite{ze}. Both methods give consistent results and 
$\varepsilon_{\rm GEC} \nolinebreak=\nolinebreak (94.0 \pm 0.2)\%$ is used in this analysis. 
The central value is the estimate from the first method, while 
the difference between the two estimates contributes to the uncertainty. 
This efficiency depends linearly on
$y_{Z}$ with about 2$\%$ variation across the full range. A weaker dependence on both the $p_{T,Z}$ and $\phi^{*}_{Z}$ 
is also observed. Corrections for these effects are made. 

\subsection{Final-state radiation}

The FSR correction is taken to be the mean of the corrections calculated with \text{\sc Herwig++}~\cite{herwig} and \text{\sc Pythia8}~\cite{pythia8}. 
The corrections are tabulated in Appendix~\ref{sec:difftab} and are on average about 2.5$\%$. 

\subsection{Unfolding detector response}
\label{sec:unfld}
To correct for detector resolution effects, an unfolding is performed (matrix $U$ of Equation~\ref{eq:cs1}) using \lhcb 
simulation and the 
\text{\sc RooUnfold}~\cite{roounfold} software package.
The momentum resolution in the simulation is calibrated to the data.
The data are then unfolded using the iterative Bayesian approach proposed in Ref.~\cite{dagostini}.
Other unfolding techniques~\cite{svd,cowan} 
give similar results.
Additionally, all unfolding methods are tested for model
dependence using underlying distributions from leading order \pythia~\cite{pythia,pythia8}, leading order \text{\sc Herwig++}~\cite{herwig}, as well as 
NLO \powheg~\cite{powheg, powheg1, powheg2} showered with both \pythia and \herwig using the
\powheg matching scheme. 
The correction is on average about 2$\%$ as a function of $p_{T,Z}$, while 
it is significantly less as a function of $\phi^{*}_{Z}$.
Only the $p_{T,Z}$ and $\phi^{*}_{Z}$ distributions are unfolded. Since $y_{Z}$ is well measured, no 
unfolding is performed and $U$ is the identity matrix in this case. 

\subsection{Systematic uncertainties}
\label{sec:syst}

Sources of systematic uncertainty
and their effect on the total cross-section measurement are summarised in Table~\ref{tab:syst}.\footnote{Many of 
the systematic uncertainties quoted
here have a statistical component. The statistical uncertainty quoted on the measurement
is due to the number of observed \PZ candidates. In the case of unfolded measurements, the
statistical uncertainty is provided by the covariance matrix returned by \text{\sc RooUnfold}.}
The measured cross-sections as a function of $p_{T,Z}$ and $\phi^{*}_{Z}$ 
have additional systematic uncertainties
due to unfolding.
\begin{table}[h]
\caption{Contributions to the relative uncertainty on the total $\PZ$ boson cross-section.}
\label{tab:syst}
\begin{center}\begin{tabular}{lr}
    Source                           & Uncertainty ($\%$)  \\ \hline
    Statistical               & $0.39$ \\ \hline
    Trigger efficiency        & $  0.07 $          \\
    Identification efficiency & $  0.23 $         \\
    Tracking efficiency       & $  0.53 $           \\
    FSR                 & $  0.11 $                \\
    Purity              & $  0.22 $                 \\
    GEC efficiency            & $  0.26 $                  \\ \hline
    Systematic               & $0.68$                         \\ \hline
    Beam energy          & $1.25$    \\
    Luminosity          & $1.72$    \\ \hline
    Total                & $2.27$
  \end{tabular}\end{center}
\end{table}

The systematic uncertainty associated with the trigger, identification and tracking efficiencies is
determined by
re-evaluating all cross-sections
with the values of the individual efficiencies increased or decreased by one standard deviation. 
The full covariance matrix of the differential cross-section measurements is evaluated in this way
for each source of uncertainty separately. 
The covariance matrices for each source are added and the diagonal elements of the result determine
the total systematic uncertainty due to reconstruction efficiencies. These 
vary between
0.5 and 2.0$\%$ on the differential cross-section measurements.

The systematic uncertainty on the FSR correction is the quadratic sum of two components. The first is due
to the statistical precision of the \pythia and \text{\sc Herwig++} estimates and the second is half 
of the difference between their central values. The latter dominates, with the uncertainties on the differential
cross-sections varying between 0.3 and 3$\%$.

The systematic uncertainty on the purity is determined from the number of candidate and background events.   
In addition, an uncertainty based on the assumption that the purity
is the same for all variables and bins of the analysis is 
evaluated by comparing to cross-section measurements using
a binned purity, rather than a global one. The total uncertainties on the differential cross-section 
measurements due to variations in purity are typically less than 1$\%$.

The GEC efficiency is determined in each bin of $y_{Z}$, $p_{T,Z}$ and $\phi^{*}_{Z}$. The systematic uncertainty is the sum in quadrature of a 
component due to the available sample size in each bin and a component due to the 0.2$\%$ uncertainty on the integrated number, as determined in 
Section~\ref{sec:gec}. 
This varies between 0.4 and 4$\%$ across the differential measurements.

The systematic uncertainty due to unfolding is estimated by the differences between the differential cross-sections using Bayesian and matrix 
inversion unfolding techniques. The typical size is $1.5\%$.

The measurement is specified at centre-of-mass energy $\sqrt{s}$ = 7 \tev.
The beam energy, and consequently the centre-of-mass energy, is known to 0.65$\%$~\cite{BEAM}. 
The sensitivity of the cross-section 
to the centre-of-mass energy is studied using \dynnlo~\cite{dynnlo} and a systematic uncertainty of 1.25$\%$ is assigned.  

\section{Results} 
\label{sec:results}
\subsection{\emph{\textbf{Z}} boson production cross-section}
\label{sec:zresults}
\begin{figure}[tbh]
  \begin{center}
    
\includegraphics[width=0.75\linewidth]{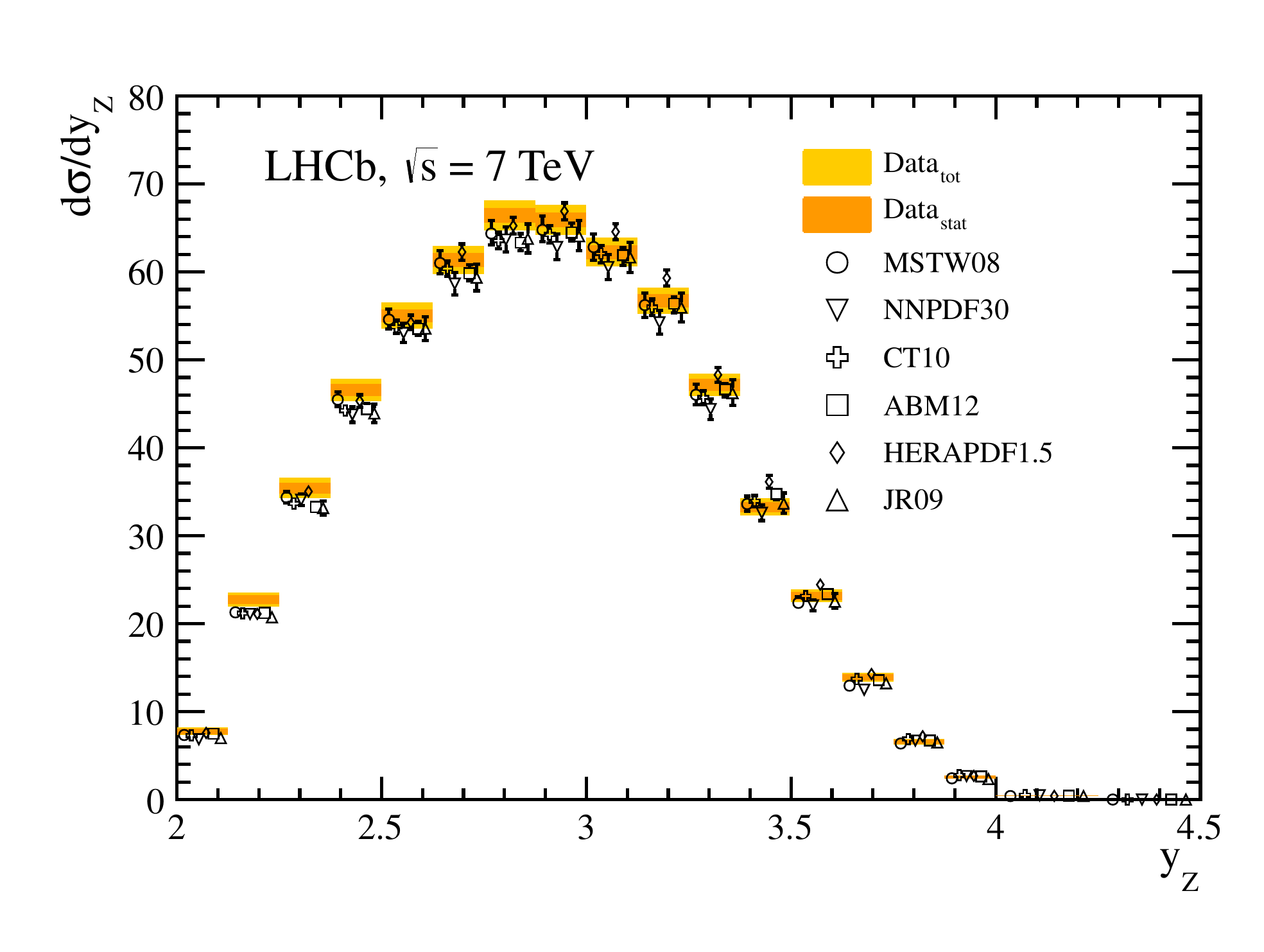}\put(-40,215){(a)} \\
    \includegraphics[width=0.75\linewidth]{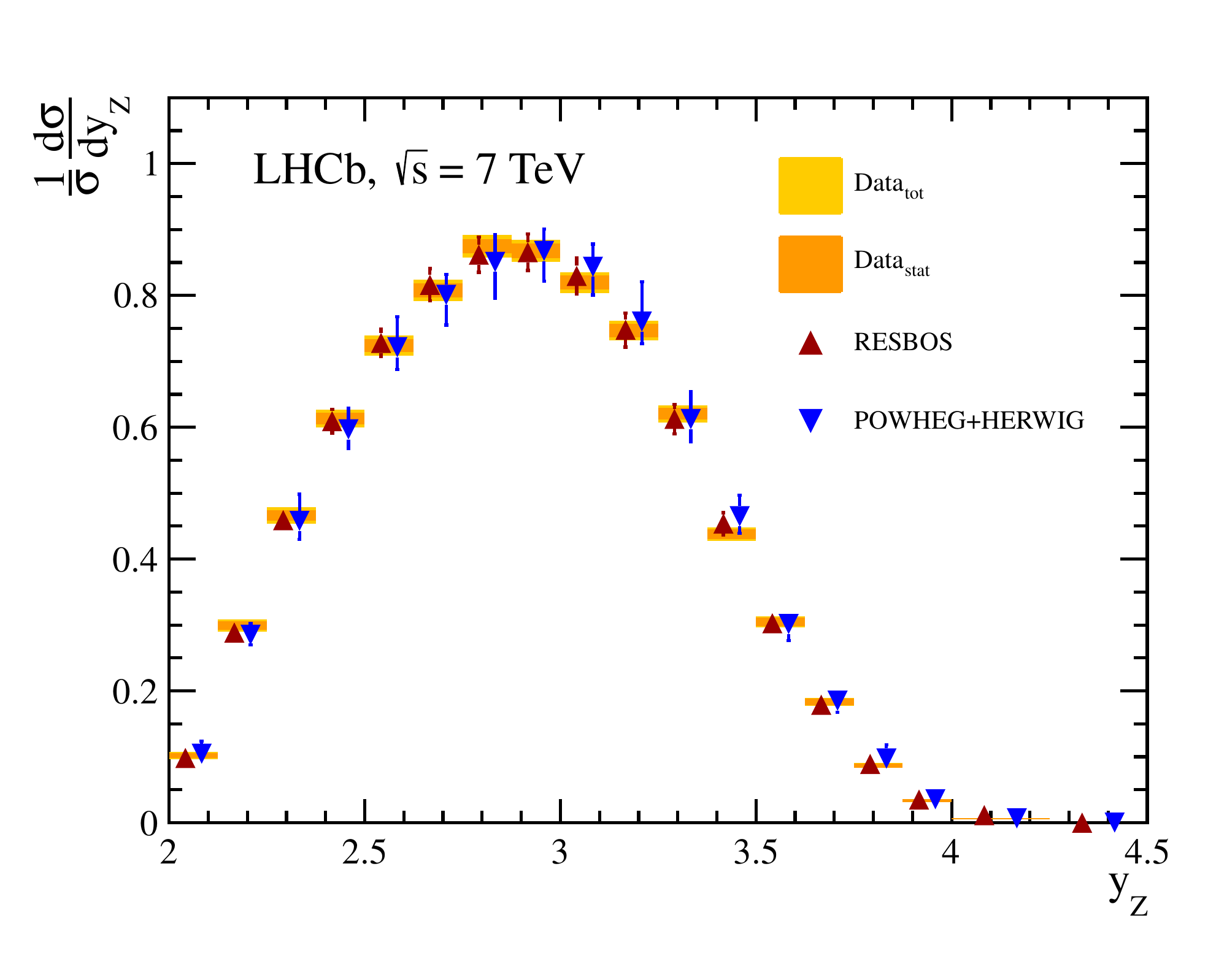}\put(-40,215){(b)}
  \end{center}
  \vspace{-1cm}\caption{
   (a) Differential cross-section as a function of $y_{Z}$ 
   compared with the prediction of \fewz configured with various PDF sets. 
   Different predictions are displaced horizontally for visibility.
   (b) Normalised differential cross-section as a function of $y_{Z}$ 
compared to the predictions of \resbos and \powheg + \herwig.
   The shaded (yellow) bands indicate the statistical and total 
uncertainties on the measurements, which are symmetric about the central value.
  }
  \label{fig:diffxsecY}
\end{figure}

\begin{figure}[tbh]
  \begin{center}
    \includegraphics[width=0.75\linewidth]{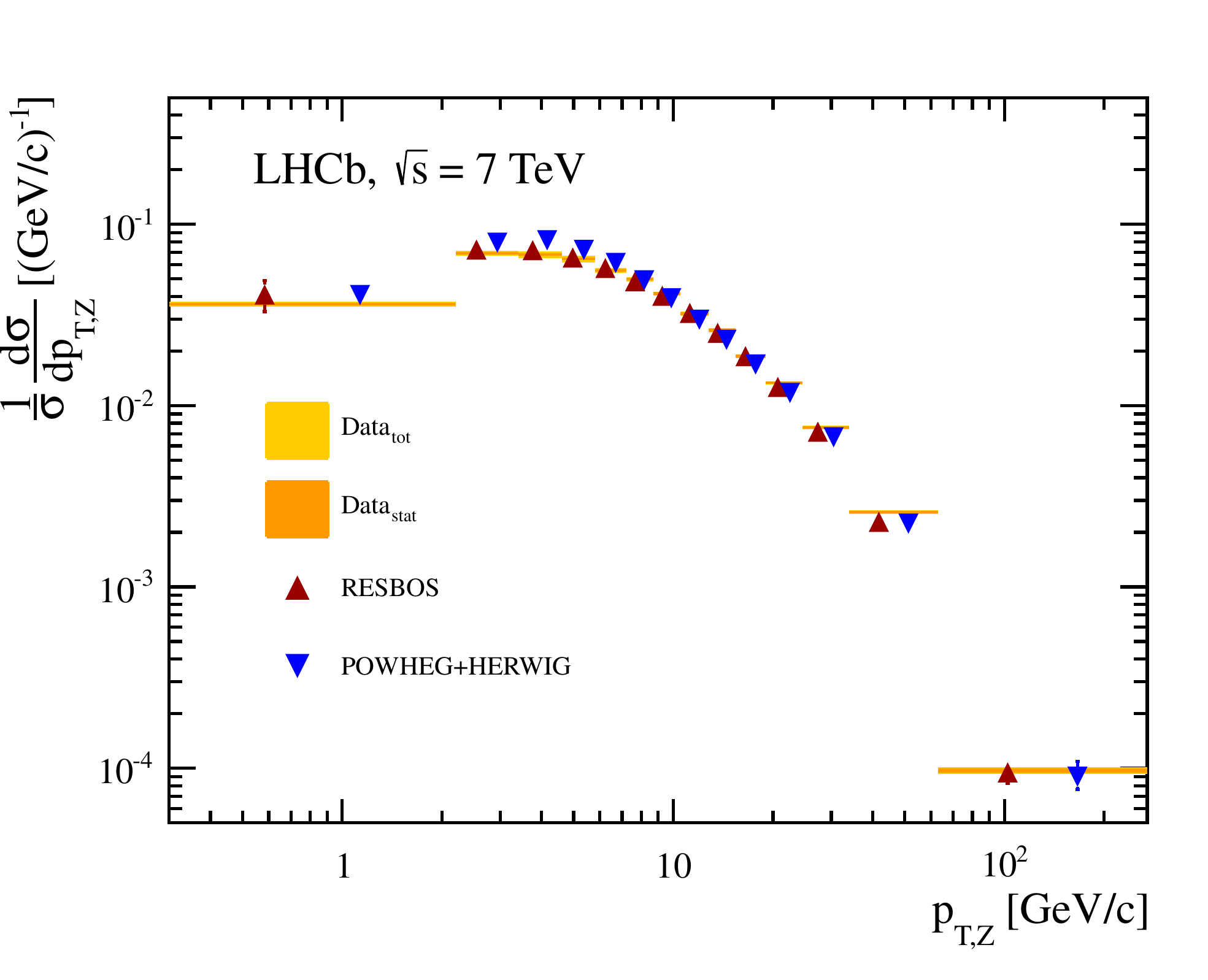}\put(-40,215){(a)} \\
    \includegraphics[width=0.75\linewidth]{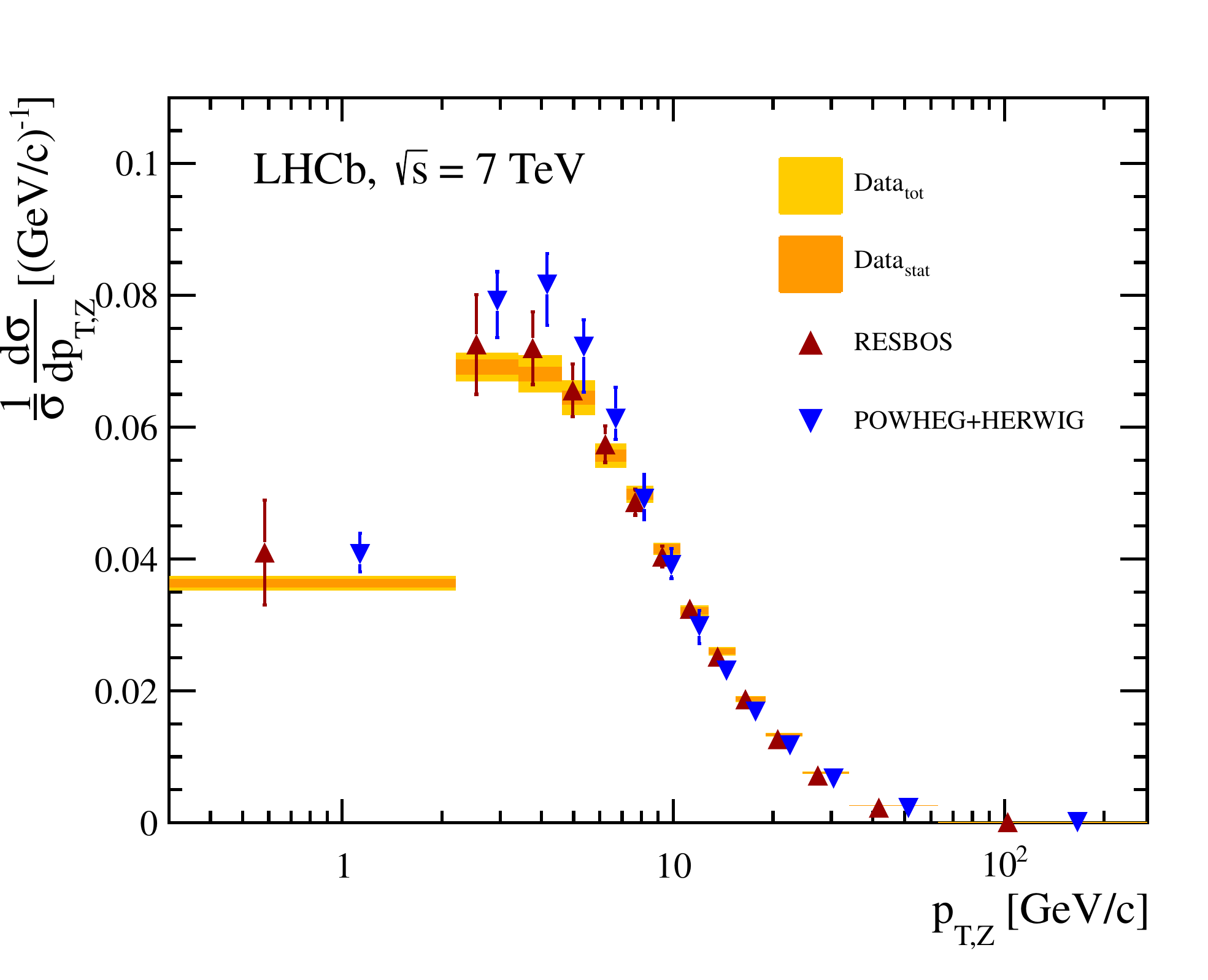}\put(-40,215){(b)}
  \end{center}
  \vspace{-1cm}\caption{
   Normalised differential cross-section as a function of $p_{T,Z}$ on (a) logarithmic 
   and (b) linear scales.
   The shaded (yellow) bands indicate the statistical and total 
uncertainties on the measurements, which are symmetric about the central
value. The measurements are compared to the predictions of \resbos and \powheg + \herwig.
  }
  \label{fig:diffxsecPT}
\end{figure}

\begin{figure}[tbh]
  \begin{center}
    \includegraphics[width=0.75\linewidth]{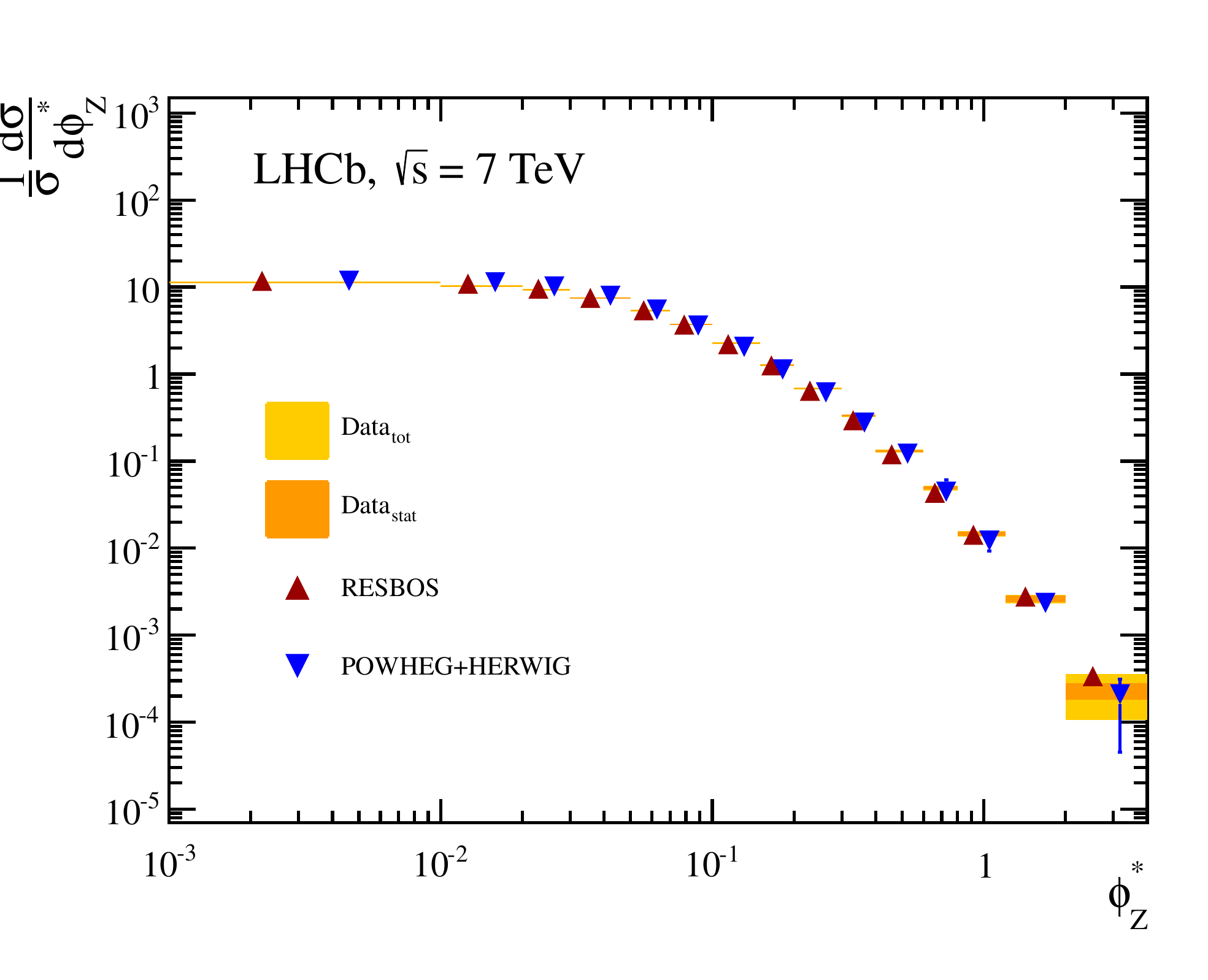}\put(-40,215){(a)} \\
    \includegraphics[width=0.75\linewidth]{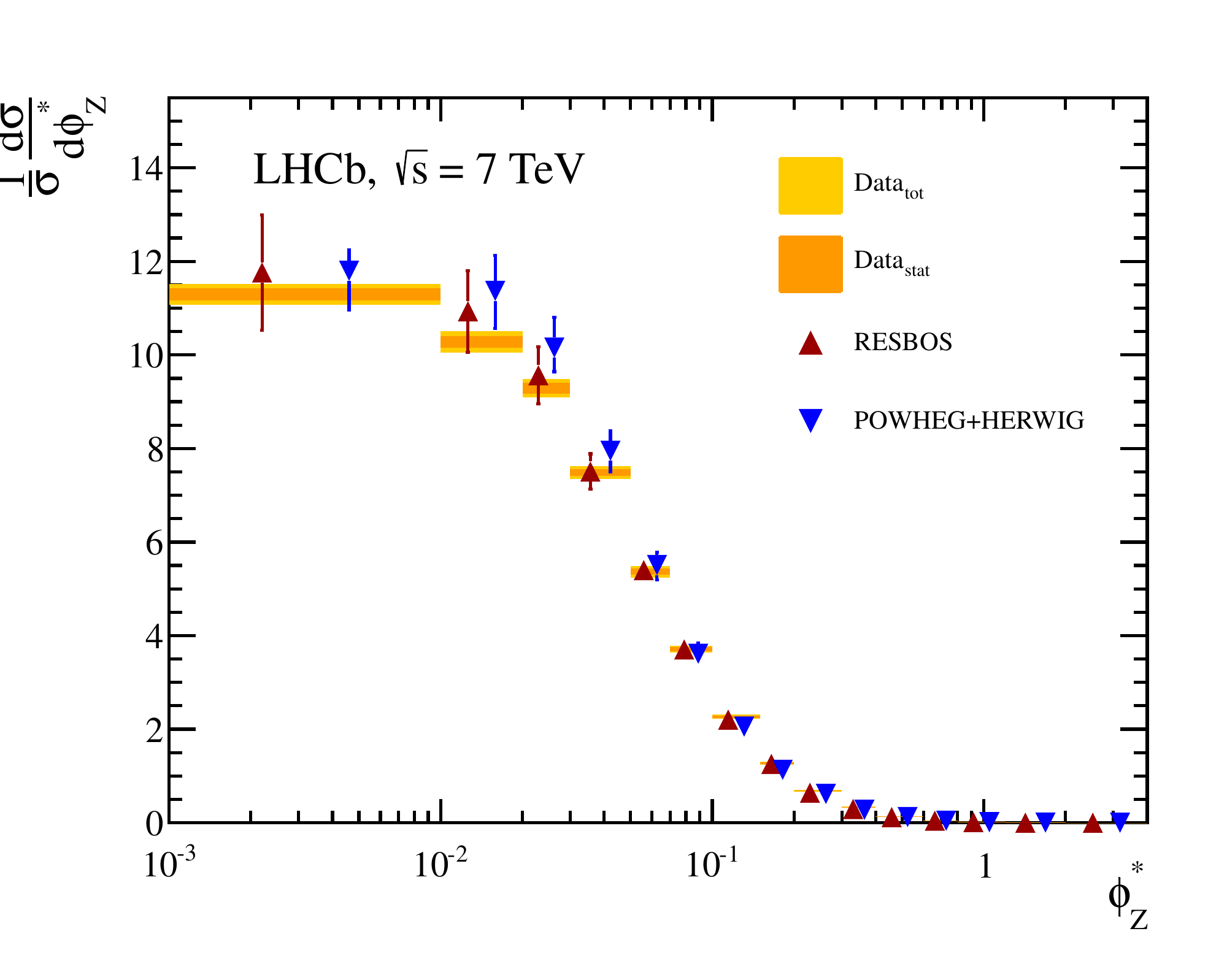}\put(-40,215){(b)}
  \end{center}
  \vspace{-1cm}\caption{
   Normalised differential cross-section as a function of $\phi^{*}_{Z}$ on (a) logarithmic
   and (b) linear scales.
   The shaded (yellow) bands indicate the statistical and total 
uncertainties on the measurements, which are symmetric about the central 
value. The measurements are compared to the predictions of \resbos and \powheg + \herwig. 
  }
  \label{fig:diffxsecPHI}
\end{figure}

\begin{figure}[tbh]
  \begin{center}
    \includegraphics[width=0.75\linewidth]{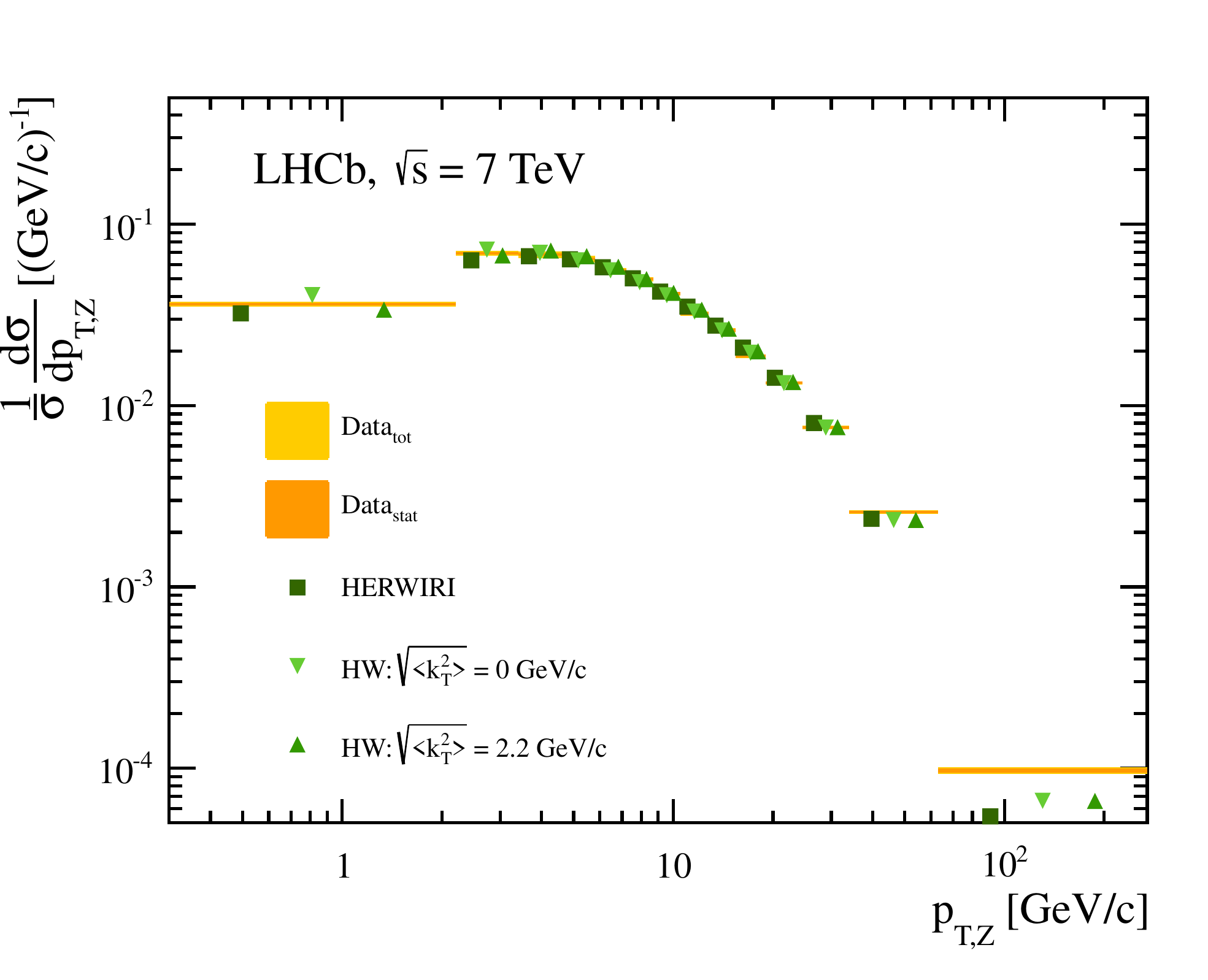}\put(-40,215){(a)} \\
    \includegraphics[width=0.75\linewidth]{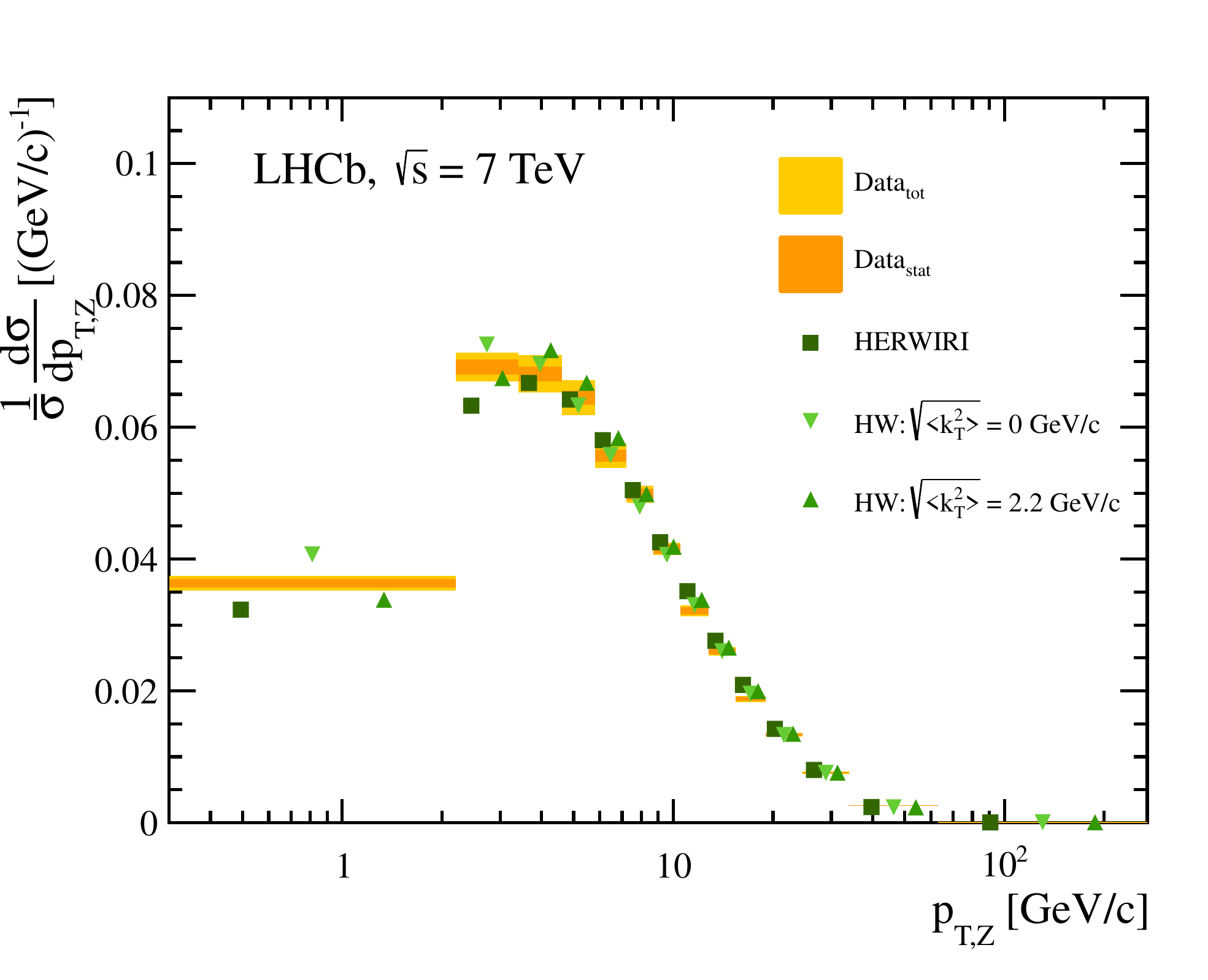}\put(-40,215){(b)}
  \end{center}
  \vspace{-1cm}\caption{
   Normalised differential cross-section as a function of $p_{T,Z}$ on (a) logarithmic
   and (b) linear scales.
The shaded (yellow) bands indicate the statistical 
and total uncertainties on the measurements, which are symmetric about the 
central value. The measurements are compared to \mc@nlo + \herwig (HW) and 
\mc@nlo + \herwiri (HERWIRI). \herwig is configured with two choices of the 
root mean-square-deviation of the intrinsic $k_{T}$ distribution, 0 and 2.2 
\gevc.
  }
  \label{fig:diffxsecHERWIRI_PT}
\end{figure}

\begin{figure}[tbh]
  \begin{center}
    \includegraphics[width=0.75\linewidth]{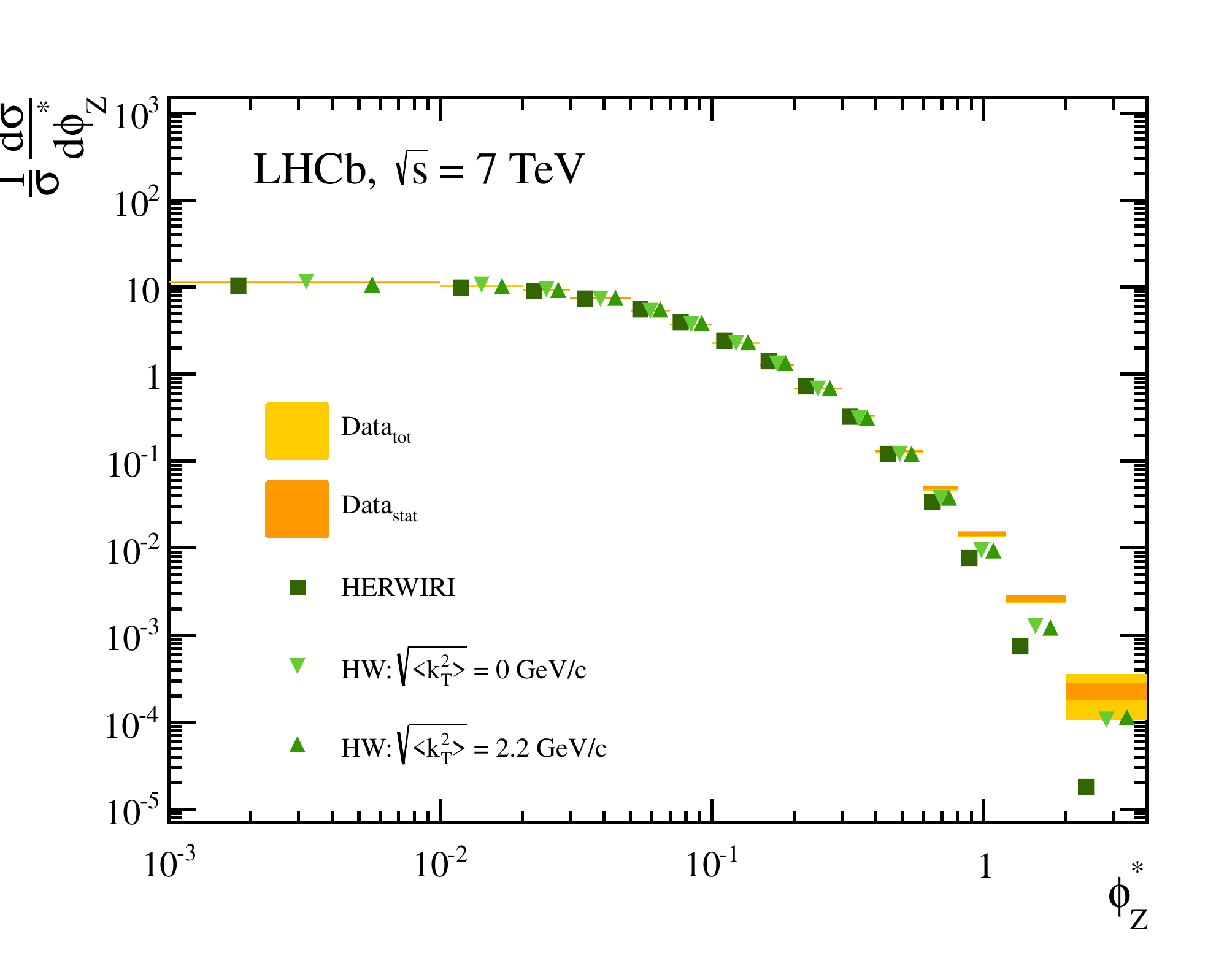}\put(-40,215){(a)} \\
    \includegraphics[width=0.75\linewidth]{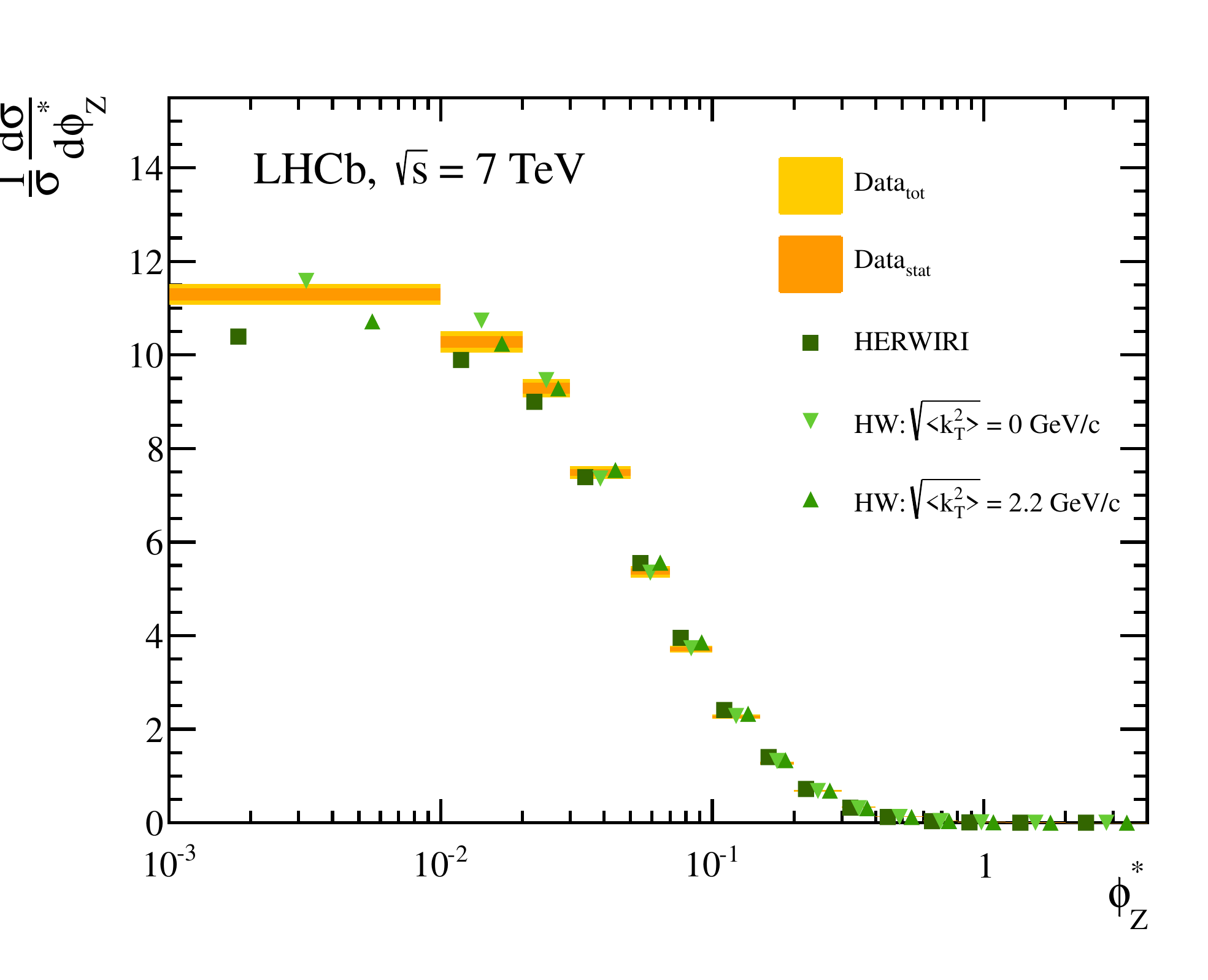}\put(-40,215){(b)}
  \end{center}
  \vspace{-1cm}\caption{
   Normalised differential cross-section as a function of $\phi^{*}_{Z}$ on (a) logarithmic
   and (b) linear scales.
The shaded (yellow) bands indicate the statistical        
and total uncertainties on the measurements, which are symmetric about the
central value. The measurements are compared to \mc@nlo + \herwig (HW) and 
\mc@nlo + \herwiri (HERWIRI). \herwig is configured with two choices of the 
root mean-square-deviation of the intrinsic $k_{T}$ distribution, 0 and 2.2 
\gevc.
  }
  \label{fig:diffxsecHERWIRI_PHI}
\end{figure}

The measured rapidity distribution as shown in Fig.~\ref{fig:diffxsecY} is 
compared to the prediction from \fewz~\cite{fewz,fewzold} with six different PDF sets.   
To compare the shapes of the differential cross-sections, measurements and
predictions are normalised to the total fiducial cross-section.
The normalised differential cross-sections are shown in 
Figs.~\ref{fig:diffxsecY},~\ref{fig:diffxsecPT} and~\ref{fig:diffxsecPHI}.
The measurements are compared to the predictions from \resbos~\cite{resbos1,resbos2,resbos3} and \powheg~\cite{powheg} where events
are interfaced with a parton shower that is simulated using \herwig~\cite{GEN-HERWIG1,GEN-HERWIG2}.
The $p_{T,Z}$ and $\phi^{*}_{Z}$ distributions are well described by \resbos and \powheg,
with the central values overestimating the data slightly at low $\phi^{*}_{Z}$ and underestimating slightly at high $\phi^{*}_{Z}$.
Comparisons to \mc@nlo + \herwiri and \mc@nlo + \herwig are shown in 
Figs.~\ref{fig:diffxsecHERWIRI_PT} and~\ref{fig:diffxsecHERWIRI_PHI}.
Here \herwig is configured with the root mean-square-deviation of the 
intrinsic $k_{T}$ distribution set to 0\gevc in
one instance and 2.2\gevc in another. The predictions straddle the measurement at low $p_{T,Z}$ and $\phi^{*}_{Z}$.
The high $p_{T,Z}$ and $\phi^{*}_{Z}$ tails are underestimated.

The total inclusive cross-section for $\PZ\rightarrow\mu^{+}\mu^{-}$ production for muons with $\pt \nolinebreak>\nolinebreak 20\nolinebreak\gevc$ in the
pseudorapidity region $2.0 < \eta < 4.5$ and the dimuon invariant mass range 
$60<M_{\mu^{+}\mu^{-}}<120\gevcc$ is measured to be
\begin{equation*}
 \sigma_{\PZ\rightarrow\mu^{+}\mu^{-}} = (76.0 \pm 0.3 \pm 0.5 \pm 1.0 \pm 1.3) \, \text{pb},
\end{equation*}
where the first uncertainty is statistical, the second systematic, the third is due to the beam energy
and the fourth is due to the luminosity.
The upper plot of Fig.~\ref{fig:totcs} shows agreement between this measurement and NNLO predictions given by \fewz
configured with various PDF sets.
The measurement also agrees with the measurements of the \PZ boson production cross-section performed in the
electron~\cite{ze} and tau lepton~\cite{ztau} channels but with a significantly smaller uncertainty.
All binned cross-sections are detailed in Tables~\ref{tab:csZY},~\ref{tab:csZPT} and~\ref{tab:csZPHI} of Appendix~\ref{sec:difftab}.

\subsection{Ratios of electroweak boson production cross-sections }
\label{sec:wzcomb}

The cross-section ratios   
are defined for muons with $\pt>20\gevc$, $2.0<\eta<4.5$ and, in the case of the \PZ boson cross-section, a dimuon
invariant mass between 60 and 120$\nolinebreak\gevcc$. 
The ratio of \PW boson to \PZ boson production
is defined as
\begin{equation}
R_{\it \PW\PZ} =
\frac{\sigma_{W^{+}\rightarrow\mu^{+}\nu_{\mu}}+\sigma_{W^{-}\rightarrow\mu^{-}\bar{\nu}_{\mu}}}{\sigma_{Z\rightarrow\mu^{+}\mu^{-}}}.
\end{equation}
The separate ratios of $W^{+}$ and $W^{-}$ to \PZ boson production cross-sections are defined as
\begin{equation}
R_{\it \PW^{\pm}\PZ} =
\frac{\sigma_{W^{\pm}\rightarrow\mu^{\pm}\nu_{\mu}}}{\sigma_{Z\rightarrow\mu^{+}\mu^{-}}},
\end{equation}
while the  \PW boson cross-section ratio is defined as
\begin{equation}
R_{\it \PW} = \frac{ \sigma_{W^{+}\rightarrow\mu^{+}\nu_{\mu}} } { \sigma_{W^{-}\rightarrow\mu^{-}\bar{\nu}_{\mu}}}.
\end{equation}

Many sources of systematic uncertainty cancel or are reduced in the ratios. As the data sets are identical, the largest 
single source of uncertainty 
on the individual cross-sections, due to the luminosity determination, is removed. The trigger used to select both samples
is identical and most of the uncertainty on the determination of the trigger efficiency cancels. In particular, the GEC
is common to both the \PW and \PZ boson 
analyses and it is 
expected that the size of the efficiency correction is similar for \PW and \PZ events. 
Cross-checks in data
and simulation support this assumption with a precision of approximately 0.3$\%$, which is included as a systematic 
uncertainty. The GEC efficiency was determined for the previous \PW boson 
measurement~\cite{wmu} to be 
$(95.9\pm1.1)\%$, whereas an improved precision of $(94.0\pm0.2)\%$ is obtained in the current analysis. Consequently, the \PW boson cross-section
results are updated to benefit from the more precise value. These results are listed in 
Tables~\ref{tab:csWETA},~\ref{tab:csrw} and~\ref{tab:csaw} 
of Appendix~\ref{sec:difftab}, along with the muon charge ratios
and asymmetries, and supersede those in Ref.~\cite{wmu}. 
The uncertainties on the tracking and muon identification partially cancel in the ratios of \PW 
and \PZ bosons. The uncertainty on the $W^{+}(W^{-})$ cross-section due to beam energy is 1.06(0.91)$\%$ and most of this uncertainty also cancels in the ratios. 
The uncertainties on the purities of the \PW and \PZ boson selections are uncorrelated.
The FSR uncertainties are taken to be uncorrelated.
The sources of uncertainty contributing to the determination of the ratios are summarised in Table~\ref{tab:wzsyst}.

\begin{table}[h]
\caption{Contributions to the relative uncertainty on the electroweak boson cross-section ratios.}

\label{tab:wzsyst}

\begin{center}\begin{tabular}{l c c c c}
    Source                           & \multicolumn{4}{c}{Uncertainty ($\%$)}  \\ \hline
                                        & $R_{\it \PW\PZ}$ & $R_{\it W^{+}\PZ}$ & $R_{\it W^{-}\PZ}$ & $R_{\it \PW}$ \\ \hline
    Statistical                         & $  0.45 $ & $  0.48 $ & $  0.50 $ & $  0.38 $           \\ \hline
    Trigger efficiency                        & $  0.15 $ & $  0.16 $ & $  0.13 $ & $  0.07 $           \\
    Identification efficiency                 & $  0.12 $ & $  0.12 $ & $  0.12 $ & $  0.03 $           \\
    Tracking efficiency                       & $  0.24 $ & $  0.23 $ & $  0.26 $ & $  0.08 $           \\
    FSR                                 & $  0.16 $ & $  0.21 $ & $  0.17 $ & $  0.21 $           \\
    Purity                              & $  0.41 $ & $  0.49 $ & $  0.55 $ & $  0.62 $           \\
    GEC efficiency                            & $  0.27 $ & $  0.28 $ & $  0.29 $ & $  0.18 $           \\ \hline
    Systematic                         & $  0.60 $ & $  0.67 $ & $  0.72 $ & $  0.69 $           \\ \hline           
    Beam energy                        & $  0.26 $ & $  0.19 $ & $  0.34 $ & $  0.15 $           \\ \hline
    Total                               & $  0.79 $ & $  0.85 $ & $  0.94 $ & $  0.80 $           \\

  \end{tabular}\end{center}
\end{table}

The dominant uncertainties on the ratios are due to the purity and the size of the sample. 
The correlation coefficients used in the 
uncertainty calculations are tabulated in Appendix~\ref{sec:correlation}.

\begin{figure}[tbh]
  \begin{center}
    \includegraphics[width=\linewidth]{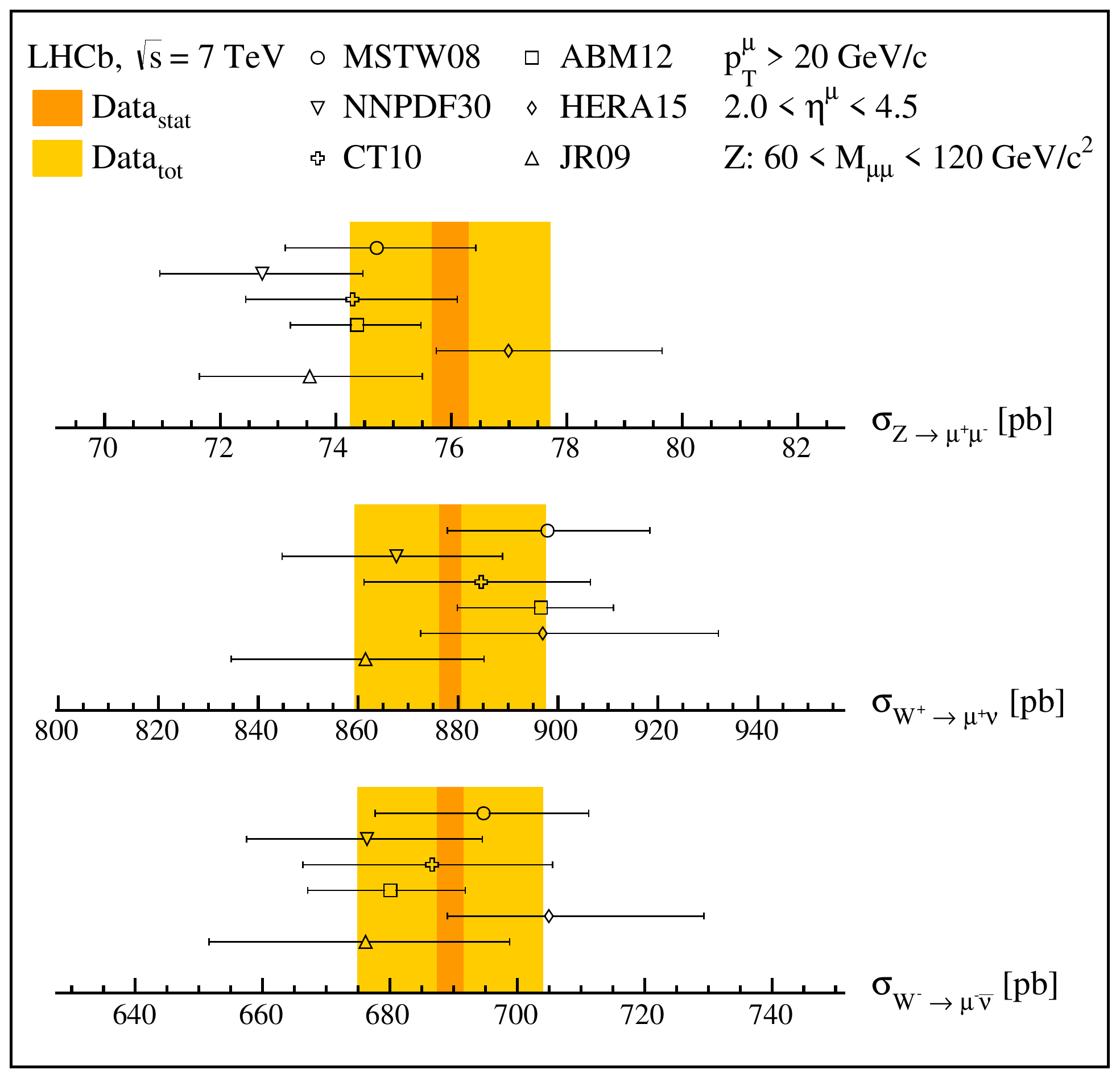}
  \end{center}
  \caption{
   \lhcb measurements of electroweak boson production cross-sections compared to NNLO pQCD as implemented by the \fewz
   generator using various PDF sets.
   The shaded (yellow) bands indicate the statistical and total 
uncertainties on the measurements, which are symmetric about the central 
value.
  }
  \label{fig:totcs}
\end{figure}

\begin{figure}[tbh]
  \begin{center}
    \includegraphics[width=\linewidth]{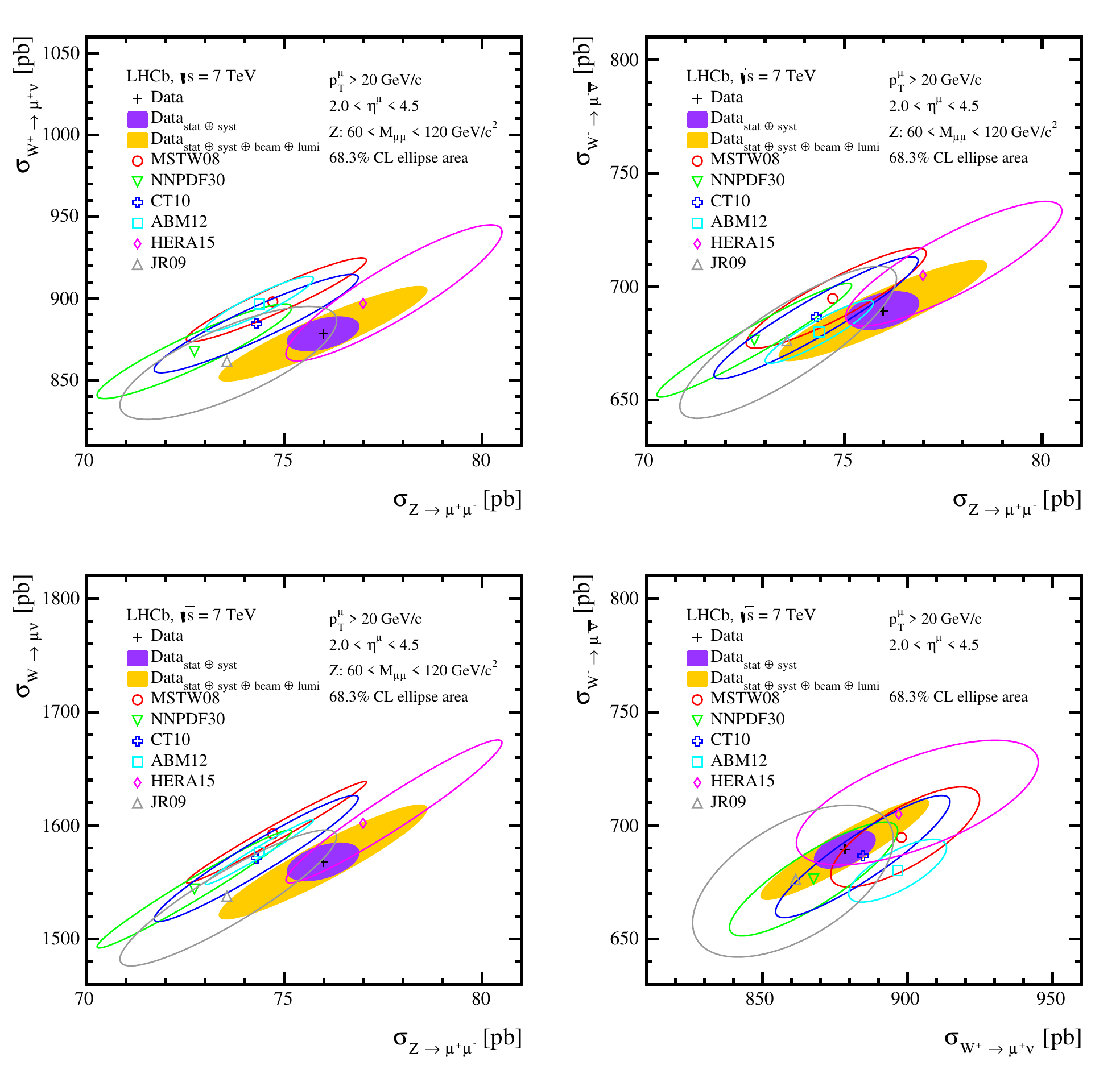}
  \end{center}
  \caption{
   Two dimensional plots of electroweak boson cross-sections compared to NNLO predictions for various
   parameterisations of the PDFs. The outer, shaded (yellow) ellipse 
corresponds to the 
total uncertainty on the measurements. The inner, shaded (purple) ellipse 
excludes the beam energy and luminosity uncertainties. The uncertainty on the theoretical predictions corresponds to the PDF uncertainty only.
   All ellipses correspond to uncertainties at 68.3$\%$ confidence level.
 }
\label{fig:ellipses}
\end{figure}

The updated $W^{+}$ boson cross-section is 
$$\sigma_{W^{+}\rightarrow\mu^{+}\nu_{\mu}} = (878.0 \pm 2.1 \pm 6.7 \pm 9.3 \pm 15.0) \hspace{0.2cm} \text{pb},$$
where the uncertainties are due to the sample size, systematic effects, the beam energy and the luminosity determination.
The updated $W^{-}$ boson cross-section is
$$\sigma_{W^{-}\rightarrow\mu^{-}\bar{\nu}_{\mu}} = (689.5 \pm 2.0 \pm 5.3 \pm 6.3 \pm 11.8) \hspace{0.2cm} \text{pb}.$$
These measurements are in good agreement with the predictions of NNLO pQCD, as shown in Fig.~\ref{fig:totcs}.
Using the \PZ boson cross-section from Section~\ref{sec:zresults}, electroweak boson cross-section measurements
and theoretical predictions, with different parameterisations of the PDFs, are compared in Fig.~\ref{fig:ellipses}, with  
contours corresponding to the 68.3$\%$ confidence level.

\begin{figure}[tbh]
  \begin{center}   
    \includegraphics[width=\linewidth]{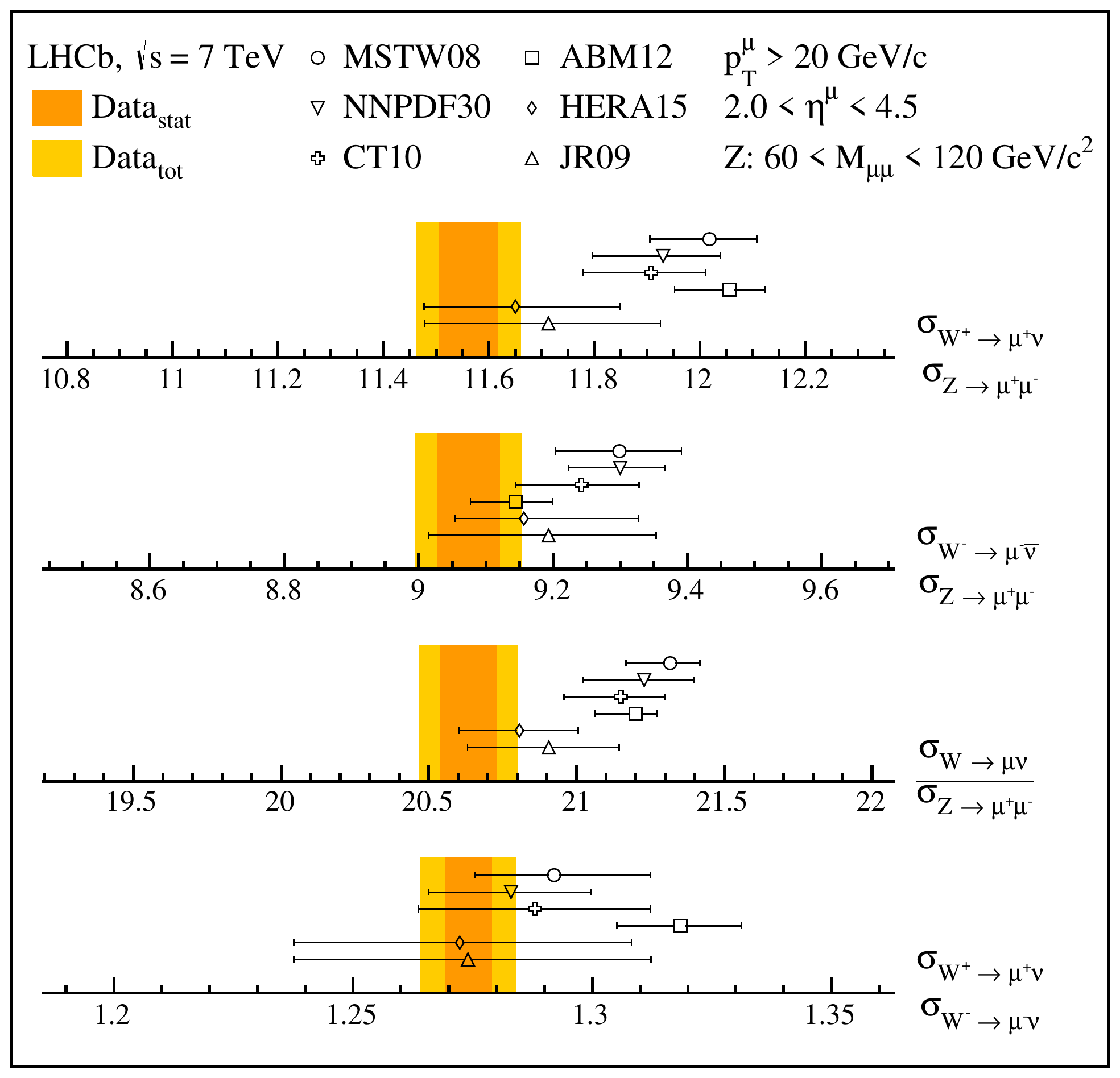}
  \end{center}
  \caption{Ratios of electroweak boson production $R_{\it W^{+}Z}$, $R_{\it W^{-}Z}$, $R_{\it WZ}$, $R_{\it W}$, compared to 
various theoretical predictions.
    The shaded (yellow) bands indicate the statistical and total 
uncertainties 
on the measurements, which are symmetric about the central value.  
  }
  \label{fig:ratiosummary}
\end{figure}

The \PW to \PZ boson cross-section ratio is measured as $$R_{\it \PW\PZ} = 20.63\pm0.09\pm0.12\pm0.05,$$
where the first uncertainty is statistical, the second is systematic and the third is due to the beam energy.
The charged \PW to \PZ boson cross-section ratios are measured as 
\begin{align*}
R_{\it W^{+}\PZ} &= 11.56\pm0.06\pm0.08\pm0.02, \\
R_{\it W^{-}\PZ} &= \phantom{0}9.07\pm0.05\pm0.07\pm0.03,  
\end{align*}
while the \PW boson cross-section ratio is measured as $$R_{\it \PW} = 1.274\pm0.005\pm0.009\pm0.002.$$
These measurements, as well as their predictions, are displayed in Fig.~\ref{fig:ratiosummary}.
For $R_{\it \PW\PZ}$ and $R_{\it W^{+}\PZ}$, the data are well described by HERA1.5 and JR09, while
the values from CT10, MSTW08, NNPDF3.0 and ABM12 are larger than those measured. 
All PDF sets show good agreement for $R_{\it W^{-}\PZ}$.
As previously reported~\cite{wmu}, all PDF sets except ABM12 show good agreement for $R_{\it \PW}$. 
The $R_{\it \PW\PZ}$ and $R_{\it \PW}$ ratios are measured with a 
fractional uncertainty of $0.8\%$, which is similar both to 
the precision due to the PDFs
on the individual theoretical predictions and to the spread between the predictions. 
Considering the spread in the different predictions, the experimental measurements are in good agreement with SM predictions and can be used 
to improve the determination of the PDFs.

\section{Conclusions} 
\label{sec:conclusions}

A measurement of the forward \PZ boson production cross-section at $\sqrt{s}$ = 7 \tev is presented, where
the \PZ bosons are reconstructed in the decay $Z\rightarrow\mu^{+}\mu^{-}$. The total cross-section 
in the fiducial range of the selection is in agreement with NNLO pQCD calculations. 
Normalised differential cross-sections as a function of $y_{Z}$, 
$\phi^{*}_{Z}$ and $p_{T,Z}$ are compared
to the predictions of various generators.
The increased precision on the determination of the event trigger efficiency motivates a re-evaluation of the
recently measured \PW boson production cross-section. These are presented here and supersede the values given in Ref.~\cite{wmu}.
Combining the \PZ boson cross-section with updated \PW boson cross-sections measured in a similar fiducial volume allows for
precision measurements of electroweak boson cross-section ratios. In particular, the \PW to \PZ boson
ratio is determined with a relative precision of $0.8\%$. The measured ratios are consistent with SM predictions
but are sensitive to the particular choice of PDF. Consequently, these results are expected to provide 
significant constraints on PDFs.




\section*{Acknowledgements}
\noindent We thank B.F.L. Ward for providing \mc@nlo + \herwig and \mc@nlo + \herwiri predictions.
We express our gratitude to our colleagues in the CERN
accelerator departments for the excellent performance of the LHC. We
thank the technical and administrative staff at the LHCb
institutes. We acknowledge support from CERN and from the national
agencies: CAPES, CNPq, FAPERJ and FINEP (Brazil); NSFC (China);
CNRS/IN2P3 (France); BMBF, DFG, HGF and MPG (Germany); INFN (Italy); 
FOM and NWO (The Netherlands); MNiSW and NCN (Poland); MEN/IFA (Romania); 
MinES and FANO (Russia); MinECo (Spain); SNSF and SER (Switzerland); 
NASU (Ukraine); STFC (United Kingdom); NSF (USA).
The Tier1 computing centres are supported by IN2P3 (France), KIT and BMBF 
(Germany), INFN (Italy), NWO and SURF (The Netherlands), PIC (Spain), GridPP 
(United Kingdom).
We are indebted to the communities behind the multiple open 
source software packages on which we depend. We are also thankful for the 
computing resources and the access to software R\&D tools provided by Yandex LLC (Russia).
Individual groups or members have received support from 
EPLANET, Marie Sk\l{}odowska-Curie Actions and ERC (European Union), 
Conseil g\'{e}n\'{e}ral de Haute-Savoie, Labex ENIGMASS and OCEVU, 
R\'{e}gion Auvergne (France), RFBR (Russia), XuntaGal and GENCAT (Spain), Royal Society and Royal
Commission for the Exhibition of 1851 (United Kingdom).


\clearpage

{\noindent\bf\Large Appendices}

\appendix

\section{Cross-sections}
\label{sec:difftab}

\begin{table}[h]
\caption{Inclusive differential cross-sections for \PZ boson production as a function of 
$y_{Z}$. Uncertainties are due to the sample size, 
systematic effects,
the beam energy and the luminosity. No candidates are observed in the 4.250 -- 4.500 bin.}
\label{tab:csZY}
\begin{center}
\begin{tabular}{c | c | c}
$y_{Z}$ & $\sigma_{\PZ}$ [pb] & $f_{\rm FSR}$\\ \hline

2.000 -- 2.125 &     0.969        $\pm$     0.039 $\pm$ 0.032 $\pm$ 0.012 $\pm$ 0.017 & 1.050$\pm$0.020\\
2.125 -- 2.250 &     2.840        $\pm$     0.063 $\pm$ 0.050 $\pm$ 0.036 $\pm$ 0.049 & 1.032$\pm$0.008\\
2.250 -- 2.375 &     4.428        $\pm$     0.077 $\pm$ 0.078 $\pm$ 0.055 $\pm$ 0.076 & 1.027$\pm$0.006\\
2.375 -- 2.500 &     5.823        $\pm$     0.088 $\pm$ 0.060 $\pm$ 0.073 $\pm$ 0.100 & 1.026$\pm$0.004\\
2.500 -- 2.625 &     6.877        $\pm$     0.095 $\pm$ 0.068 $\pm$ 0.086 $\pm$ 0.118 & 1.025$\pm$0.004\\
2.625 -- 2.750 &     7.669        $\pm$     0.100 $\pm$ 0.069 $\pm$ 0.096 $\pm$ 0.132 & 1.026$\pm$0.004\\
2.750 -- 2.875 &     8.306        $\pm$     0.104 $\pm$ 0.070 $\pm$ 0.104 $\pm$ 0.143 & 1.026$\pm$0.003\\
2.875 -- 3.000 &     8.241        $\pm$     0.103 $\pm$ 0.066 $\pm$ 0.103 $\pm$ 0.142 & 1.025$\pm$0.003\\
3.000 -- 3.125 &     7.783        $\pm$     0.099 $\pm$ 0.059 $\pm$ 0.097 $\pm$ 0.134 & 1.026$\pm$0.003\\
3.125 -- 3.250 &     7.094        $\pm$     0.096 $\pm$ 0.058 $\pm$ 0.089 $\pm$ 0.122 & 1.028$\pm$0.004\\
3.250 -- 3.375 &     5.894        $\pm$     0.087 $\pm$ 0.049 $\pm$ 0.074 $\pm$ 0.101 & 1.026$\pm$0.004\\
3.375 -- 3.500 &     4.160        $\pm$     0.073 $\pm$ 0.041 $\pm$ 0.052 $\pm$ 0.072 & 1.027$\pm$0.005\\
3.500 -- 3.625 &     2.896        $\pm$     0.061 $\pm$ 0.030 $\pm$ 0.036 $\pm$ 0.050 & 1.026$\pm$0.005\\
3.625 -- 3.750 &     1.741        $\pm$     0.047 $\pm$ 0.023 $\pm$ 0.022 $\pm$ 0.030 & 1.021$\pm$0.007\\
3.750 -- 3.875 &     0.825        $\pm$     0.032 $\pm$ 0.014 $\pm$ 0.010 $\pm$ 0.014 & 1.025$\pm$0.010\\
3.875 -- 4.000 &     0.321        $\pm$     0.020 $\pm$ 0.008 $\pm$ 0.004 $\pm$ 0.006 & 1.011$\pm$0.015\\
4.000 -- 4.250 &     0.115        $\pm$     0.013 $\pm$ 0.006 $\pm$ 0.001 $\pm$ 0.002 & 1.018$\pm$0.033\\
4.250 -- 4.500 &     $-$ & $-$\\

\end{tabular}
\end{center}
\end{table}

\begin{table}[h]
\caption{Inclusive differential cross-sections for \PZ boson production as a function 
of $p_{T,Z}$. Uncertainties are due to the sample 
size, systematic effects,
the beam energy and the luminosity.}
\label{tab:csZPT}
\begin{center}
\begin{tabular}{c | c | c}
 $p_{T,Z} [\gevc]$ & $\sigma_{\PZ}$ [pb] & $f_{\rm FSR}$\\ \hline

$\phantom{0}$0.0 -- $\phantom{00}$2.2 & 6.454           $\pm$     0.105$\pm$ 0.129 $\pm$ 0.081 $\pm$ 0.111 & 1.090$\pm$0.006\\
$\phantom{0}$2.2 -- $\phantom{00}$3.4 & 6.520           $\pm$     0.106$\pm$ 0.150 $\pm$ 0.081 $\pm$ 0.112 & 1.080$\pm$0.004\\
$\phantom{0}$3.4 -- $\phantom{00}$4.6 & 6.209           $\pm$     0.102$\pm$ 0.221 $\pm$ 0.078 $\pm$ 0.107 & 1.063$\pm$0.004\\
$\phantom{0}$4.6 -- $\phantom{00}$5.8 & 5.868           $\pm$     0.099$\pm$ 0.208 $\pm$ 0.073 $\pm$ 0.101 & 1.049$\pm$0.004\\
$\phantom{0}$5.8 -- $\phantom{00}$7.2 & 5.749           $\pm$     0.098$\pm$ 0.154 $\pm$ 0.072 $\pm$ 0.099 & 1.034$\pm$0.004\\
$\phantom{0}$7.2 -- $\phantom{00}$8.7 & 5.607           $\pm$     0.098$\pm$ 0.083 $\pm$ 0.070 $\pm$ 0.096 & 1.021$\pm$0.004\\
$\phantom{0}$8.7 -- $\phantom{0}$10.5 & 5.637           $\pm$     0.098$\pm$ 0.054 $\pm$ 0.070 $\pm$ 0.097 & 1.002$\pm$0.004\\
            10.5 -- $\phantom{0}$12.8 & 5.524           $\pm$     0.096$\pm$ 0.081 $\pm$ 0.069 $\pm$ 0.095 & 0.996$\pm$0.004\\
            12.8 -- $\phantom{0}$15.4 & 5.158           $\pm$     0.092$\pm$ 0.067 $\pm$ 0.064 $\pm$ 0.089 & 0.984$\pm$0.005\\
            15.4 -- $\phantom{0}$19.0 & 4.963           $\pm$     0.087$\pm$ 0.053 $\pm$ 0.062 $\pm$ 0.085 & 0.978$\pm$0.005\\
            19.0 -- $\phantom{0}$24.5 & 5.517           $\pm$     0.088$\pm$ 0.055 $\pm$ 0.069 $\pm$ 0.095 & 0.985$\pm$0.004\\
            24.5 -- $\phantom{0}$34.0 & 5.465           $\pm$     0.085$\pm$ 0.067 $\pm$ 0.068 $\pm$ 0.094 & 1.013$\pm$0.004\\
            34.0 -- $\phantom{0}$63.0 & 5.789           $\pm$     0.085$\pm$ 0.076 $\pm$ 0.072 $\pm$ 0.100 & 1.038$\pm$0.004\\
                        63.0 -- 270.0 & 1.516           $\pm$     0.043$\pm$ 0.044 $\pm$ 0.019 $\pm$ 0.026 & 1.060$\pm$0.007\\

\end{tabular}
\end{center}
\end{table}

\begin{table}[h]
\caption{Inclusive differential cross-sections for \PZ boson production as a function 
of $\phi^{*}_{Z}$. Uncertainties are due to the sample 
size, systematic effects, the beam energy and the luminosity.}
\label{tab:csZPHI}
\begin{center}
\begin{tabular}{c | c | c}
$\phi^{*}_{Z}$ & $\sigma_{\PZ}$ [pb] & $f_{\rm FSR}$\\ \hline

0.00 -- 0.01 & $\phantom{0}$8.549           $\pm$ 0.099 $\pm$ 0.088 $\pm$ 0.107 $\pm$ 0.147 & 1.034$\pm$0.004\\
0.01 -- 0.02 & $\phantom{0}$7.805           $\pm$ 0.096 $\pm$ 0.106 $\pm$ 0.098 $\pm$ 0.134 & 1.035$\pm$0.003\\
0.02 -- 0.03 & $\phantom{0}$7.051           $\pm$ 0.091 $\pm$ 0.083 $\pm$ 0.088 $\pm$ 0.121 & 1.034$\pm$0.004\\
0.03 -- 0.05 &             11.362           $\pm$ 0.114 $\pm$ 0.108 $\pm$ 0.142 $\pm$ 0.195 & 1.029$\pm$0.003\\
0.05 -- 0.07 & $\phantom{0}$8.124           $\pm$ 0.097 $\pm$ 0.120 $\pm$ 0.102 $\pm$ 0.140 & 1.026$\pm$0.003\\
0.07 -- 0.10 & $\phantom{0}$8.436           $\pm$ 0.097 $\pm$ 0.074 $\pm$ 0.105 $\pm$ 0.145 & 1.021$\pm$0.003\\
0.10 -- 0.15 & $\phantom{0}$8.611           $\pm$ 0.098 $\pm$ 0.131 $\pm$ 0.108 $\pm$ 0.148 & 1.020$\pm$0.003\\
0.15 -- 0.20 & $\phantom{0}$4.819           $\pm$ 0.073 $\pm$ 0.092 $\pm$ 0.060 $\pm$ 0.083 & 1.018$\pm$0.004\\
0.20 -- 0.30 & $\phantom{0}$5.206           $\pm$ 0.076 $\pm$ 0.058 $\pm$ 0.065 $\pm$ 0.090 & 1.019$\pm$0.004\\
0.30 -- 0.40 & $\phantom{0}$2.541           $\pm$ 0.054 $\pm$ 0.051 $\pm$ 0.032 $\pm$ 0.044 & 1.022$\pm$0.006\\
0.40 -- 0.60 & $\phantom{0}$2.018           $\pm$ 0.048 $\pm$ 0.060 $\pm$ 0.025 $\pm$ 0.035 & 1.024$\pm$0.007\\
0.60 -- 0.80 & $\phantom{0}$0.755           $\pm$ 0.029 $\pm$ 0.035 $\pm$ 0.009 $\pm$ 0.013 & 1.029$\pm$0.011\\
0.80 -- 1.20 & $\phantom{0}$0.457           $\pm$ 0.023 $\pm$ 0.018 $\pm$ 0.006 $\pm$ 0.008 & 1.025$\pm$0.014\\
1.20 -- 2.00 & $\phantom{0}$0.166           $\pm$ 0.014 $\pm$ 0.011 $\pm$ 0.002 $\pm$ 0.003 & 1.030$\pm$0.023\\
2.00 -- 4.00 & $\phantom{0}$0.045           $\pm$ 0.008 $\pm$ 0.017 $\pm$ 0.001 $\pm$ 0.001 & 1.031$\pm$0.041\\

\end{tabular}
\end{center}
\end{table}

\begin{sidewaystable}[h]
\caption{Inclusive differential cross-sections for \Wp (left) and \Wm (right) boson 
production as a function of muon $\eta$. Uncertainties 
are due to the sample size, systematic effects, the beam energy
and the luminosity. These supersede the results in Ref.~\cite{wmu}.}
\label{tab:csWETA}
\begin{center}
\begin{tabular}{c|c|c|c|c}

$\eta^{\mu}$ & $\sigma_{\Wp}$ [pb] & $f_{\rm FSR}^{\Wp}$ & $\sigma_{\Wm}$ [pb] & $f_{\rm FSR}^{\Wm}$ \\ \hline
2.00 -- 2.25 & $192.2 \pm 1.2 \pm 3.5 \pm 2.0 \pm 3.3$ & $1.016 \pm 0.004$ & $111.1 \pm 0.9 \pm 2.1 \pm 1.0 \pm 1.9$ & $1.019 \pm 0.003$ \\
2.25 -- 2.50 & $178.8 \pm 0.9 \pm 3.1 \pm 1.9 \pm 3.1$ & $1.018 \pm 0.004$ & $104.9 \pm 0.7 \pm 1.9 \pm 1.0 \pm 1.8$ & $1.015 \pm 0.003$ \\
2.50 -- 2.75 & $154.3 \pm 0.8 \pm 2.1 \pm 1.6 \pm 2.6$ & $1.025 \pm 0.005$ & $\phantom{0}96.1 \pm 0.7 \pm 1.3 \pm 0.9 \pm 1.6$ & $1.010 \pm 0.003$ \\
2.75 -- 3.00 & $122.8 \pm 0.7 \pm 1.6 \pm 1.3 \pm 2.1$ & $1.015 \pm 0.004$ & $\phantom{0}88.4 \pm 0.7 \pm 1.5 \pm 0.8 \pm 1.5$ & $1.007 \pm 0.002$ \\
3.00 -- 3.25 & $\phantom{0}94.3 \pm 0.6 \pm 1.3 \pm 1.0 \pm 1.6$ & $1.021 \pm 0.005$ & $\phantom{0}80.6 \pm 0.6 \pm 1.4 \pm 0.7 \pm 1.4$ & $1.009 \pm 0.003$ \\
3.25 -- 3.50 & $\phantom{0}61.6 \pm 0.5 \pm 0.9 \pm 0.7 \pm 1.1$ & $1.015 \pm 0.005$ & $\phantom{0}68.6 \pm 0.6 \pm 1.5 \pm 0.6 \pm 1.2$ & $1.017 \pm 0.005$ \\
3.50 -- 4.00 & $\phantom{0}60.0 \pm 0.5 \pm 0.7 \pm 0.6 \pm 1.0$ & $1.024 \pm 0.005$ & $\phantom{0}95.9 \pm 0.7 \pm 1.2 \pm 0.9 \pm 1.6$ & $1.012 \pm 0.005$ \\
4.00 -- 4.50 & $\phantom{0}14.3 \pm 0.4 \pm 0.4 \pm 0.2 \pm 0.2$ & $1.021 \pm 0.005$ & $\phantom{0}43.8 \pm 0.8 \pm 1.2 \pm 0.4 \pm 0.7$ & $1.000 \pm 0.000$ \\

\end{tabular}
\end{center}
\end{sidewaystable}

\clearpage

\begin{table}[h]
\caption{\Wp to \Wm boson production cross-section ratios as a function of
muon $\eta$. Uncertainties are due to the
sample size, systematic effects and the beam energy. These supersede the results in Ref.~\cite{wmu}.}
\label{tab:csrw}
\begin{center}
\begin{tabular}{c|c}
$\eta^{\mu}$ & $R_{W}$ \\ \hline
2.00 -- 2.25 & $1.730 \pm 0.018 \pm 0.030 \pm 0.003$ \\
2.25 -- 2.50 & $1.706 \pm 0.015 \pm 0.040 \pm 0.003$ \\
2.50 -- 2.75 & $1.606 \pm 0.014 \pm 0.021 \pm 0.002$ \\
2.75 -- 3.00 & $1.388 \pm 0.013 \pm 0.024 \pm 0.002$ \\
3.00 -- 3.25 & $1.169 \pm 0.012 \pm 0.021 \pm 0.002$ \\
3.25 -- 3.50 & $0.898 \pm 0.010 \pm 0.025 \pm 0.001$ \\
3.50 -- 4.00 & $0.626 \pm 0.007 \pm 0.006 \pm 0.001$ \\
4.00 -- 4.50 & $0.328 \pm 0.011 \pm 0.011 \pm 0.000$ \\
\end{tabular}
\end{center}
\end{table}

\begin{table}[h]
\caption{Lepton charge asymmetries as a function of muon $\eta$.
Uncertainties are due to the sample size, systematic  
effects and the beam energy. These supersede the results in Ref.~\cite{wmu}.}
\label{tab:csaw}
\begin{center}
\begin{tabular}{c|c}
$\eta^{\mu}$ & $A_{\mu}$ [\%] \\ \hline
2.00 -- 2.25 & \phantom{$-$}$26.74 \pm 0.48 \pm 0.82 \pm 0.07$ \\
2.25 -- 2.50 & \phantom{$-$}$26.08 \pm 0.41 \pm 1.09 \pm 0.07$ \\
2.50 -- 2.75 & \phantom{$-$}$23.25 \pm 0.42 \pm 0.60 \pm 0.07$ \\
2.75 -- 3.00 & \phantom{$-$}$16.26 \pm 0.46 \pm 0.84 \pm 0.07$ \\
3.00 -- 3.25 & \phantom{$-0$}$7.81 \pm 0.50 \pm 0.90 \pm 0.08$ \\
3.25 -- 3.50 & \phantom{$0$}$-5.37 \pm 0.57 \pm 1.35 \pm 0.08$ \\
3.50 -- 4.00 &                      $-23.04 \pm 0.52 \pm 0.49 \pm 0.07$ \\
4.00 -- 4.50 &                      $-50.65 \pm 1.22 \pm 1.30 \pm 0.06$ \\
\end{tabular}
\end{center}
\end{table}

\clearpage

\section{Correlation Matrices}
\label{sec:correlation}

\begin{sidewaystable}
 \caption{Correlation coefficients of differential cross-section 
measurements as a function of $y_{Z}$. The beam energy and luminosity 
uncertainties, which are 
fully correlated between cross-section measurements, are excluded.}
 \resizebox{25cm}{5cm}{
  \begin{tabular}{c | c c c c c c c c c c c c c c c c c c}

$y_{\PZ}$& \scriptsize{2--2.125} & \scriptsize{2.125--2.25} & \scriptsize{2.25--2.375} & \scriptsize{2.375--2.5} & \scriptsize{2.5--2.625} & \scriptsize{2.625--2.75} & 
\scriptsize{2.75--2.875} & \scriptsize{2.875--3} & \scriptsize{3--3.125} &
\scriptsize{3.125--3.25} & \scriptsize{3.25--3.375} & \scriptsize{3.375--3.5} & \scriptsize{3.5--3.625} &
\scriptsize{3.625--3.75} &
\scriptsize{3.75--3.875} & \scriptsize{3.875--4} & \scriptsize{4--4.25}  & \scriptsize{4.25--4.5} \\ \hline

\scriptsize{2--2.125} &1& & & & & & & & & & & & & & & & & \\   
\scriptsize{2.125--2.25} &0.18& 1& & & & & & & & & & & & & & & &  \\
\scriptsize{2.25--2.375} &0.14& 0.19& 1& & & & & & & & & & & & & & &  \\
\scriptsize{2.375--2.5} &0.14& 0.19& 0.18& 1&  & & & & & & & & & & & & & \\
\scriptsize{2.5--2.625} &0.13& 0.18& 0.16& 0.19& 1& & & & & & & & & & & & & \\ 
\scriptsize{2.625--2.75} &0.12& 0.16& 0.15& 0.18& 0.18& 1& & & & & & & & & & & & \\
\scriptsize{2.75--2.875} &0.11& 0.15& 0.14& 0.17& 0.17& 0.18& 1&   & & & & & & & & & & \\
\scriptsize{2.875--3} &0.10& 0.13& 0.13& 0.16& 0.16& 0.17& 0.17& 1& & & & & & & & & &  \\
\scriptsize{3--3.125} &0.09& 0.12& 0.12& 0.14& 0.15& 0.15& 0.16& 0.16& 1&  & & & & & & & & \\  
\scriptsize{3.125--3.25} &0.08& 0.1& 0.10& 0.12& 0.13& 0.14& 0.14& 0.14& 0.14& 1&   & & & & & & & \\ 
\scriptsize{3.25--3.375} &0.06& 0.08& 0.08& 0.11& 0.11& 0.12& 0.12& 0.13& 0.13& 0.13& 1&  & & & & & & \\
\scriptsize{3.375--3.5} &0.05& 0.06& 0.06& 0.08& 0.09& 0.09& 0.10& 0.10& 0.11& 0.11& 0.11& 1&  & & & & & \\
\scriptsize{3.5--3.625} &0.04& 0.05& 0.05& 0.06& 0.07& 0.08& 0.08& 0.09& 0.09& 0.10& 0.10& 0.10& 1& & & & &  \\ 
\scriptsize{3.625--3.75} &0.03& 0.04& 0.03& 0.04& 0.05& 0.06& 0.06& 0.07& 0.07& 0.08& 0.09& 0.08& 0.08& 1&  & & & \\
\scriptsize{3.75--3.875} &0.02& 0.03& 0.02& 0.03& 0.03& 0.04& 0.04& 0.05& 0.05& 0.06& 0.06& 0.06& 0.06& 0.06& 1&  & & \\
\scriptsize{3.875--4} &0.02& 0.02& 0.02& 0.02& 0.02& 0.02& 0.02& 0.03& 0.03& 0.04& 0.04& 0.04& 0.04& 0.04& 0.03& 1& & \\
\scriptsize{4--4.25} &0.01& 0.01& 0.01& 0.01& 0.01& 0.01& 0.01& 0.02& 0.02& 0.02& 0.03& 0.03& 0.03& 0.03& 0.02& 0.02& 1& \\
\scriptsize{4.25--4.5} &$-$& $-$& $-$& $-$& $-$& $-$& $-$& $-$& $-$& $-$& $-$& $-$& $-$& $-$& $-$& $-$& $-$& $-$ \\ \hline

\label{tab:correlY}
      \end{tabular}
}
\end{sidewaystable}

\begin{sidewaystable}
 \caption{Correlation coefficients of differential cross-section 
measurements as a function of $p_{T,Z}$. The beam energy and 
luminosity uncertainties, which 
are fully correlated between cross-section measurements, are excluded.}
 \resizebox{25cm}{5cm}{
  \begin{tabular}{c | c c c c c c c c c c c c c c }

$p_{T,Z}$ [GeV/c]& \scriptsize{0.0--2.2} & \scriptsize{2.2--3.4} & \scriptsize{3.4--4.6} & \scriptsize{4.6--5.8} & \scriptsize{5.8--7.2} & \scriptsize{7.2--8.7} 
& \scriptsize{8.7--10.5} & \scriptsize{10.5--12.8} &
\scriptsize{12.8--15.4} &
\scriptsize{15.4--19} & \scriptsize{19--24.5} & \scriptsize{24.5--34} & \scriptsize{34--63} &
\scriptsize{63--270} \\ \hline

\scriptsize{0.0--2.2} &1& & & & & & & & & & & & & \\
\scriptsize{2.2--3.4} &-0.01& 1& & & & & & & & & & & & \\
\scriptsize{3.4--4.6} &0.00& 0.03& 1& & & & & & & & & & & \\
\scriptsize{4.6--5.8} &0.04& 0.00& 0.02& 1& & & & & & & & & & \\
\scriptsize{5.8--7.2} &0.05& 0.05& 0.00& 0.03& 1& & & & & & & & & \\
\scriptsize{7.2--8.7} &0.07& 0.07& 0.05& 0.00& 0.03& 1& & & & & & & & \\
\scriptsize{8.7--10.5} &0.08& 0.08& 0.06& 0.06& 0.02& 0.02& 1& & & & & & & \\
\scriptsize{10.5--12.8} &0.07& 0.06& 0.05& 0.05& 0.07& 0.04& 0.00& 1& & & & & & \\
\scriptsize{12.8--15.4} &0.07& 0.07& 0.05& 0.05& 0.06& 0.09& 0.07& -0.01& 1& & & & & \\
\scriptsize{15.4--19} &0.08& 0.08& 0.05& 0.06& 0.07& 0.10& 0.12& 0.08& -0.01& 1& & & & \\
\scriptsize{19--24.5} &0.08& 0.08& 0.06& 0.06& 0.07& 0.10& 0.12& 0.10& 0.10& 0.02& 1& & &\\
\scriptsize{24.5--34} &0.08& 0.07& 0.05& 0.06& 0.07& 0.10& 0.11& 0.09& 0.10& 0.11& 0.05& 1& &\\
\scriptsize{34--63} &0.07& 0.06& 0.05& 0.05& 0.06& 0.08& 0.09& 0.08& 0.08& 0.09& 0.10& 0.06& 1& \\
\scriptsize{63--270} &0.20& 0.20& 0.15& 0.15& 0.19& 0.26& 0.30& 0.26& 0.27& 0.30& 0.32& 0.31& 0.30& 1\\ \hline

\label{tab:correlPt}
      \end{tabular}
}
\end{sidewaystable}

\begin{sidewaystable}
 \caption{Correlation coefficients of differential cross-section 
measurements as a function of $\phi^{*}_{Z}$. The beam energy and 
luminosity uncertainties, 
which are fully correlated between cross-section measurements, are excluded.}
 \resizebox{25cm}{5cm}{
  \begin{tabular}{c | c c c c c c c c c c c c c c c }
$\phi^{*}_{\PZ}$& \scriptsize{0.00--0.01} & \scriptsize{0.01--0.02} & \scriptsize{0.02--0.03} & \scriptsize{0.03--0.05}
& \scriptsize{0.05--0.07} & \scriptsize{0.07--0.10} & \scriptsize{0.10--0.15} & \scriptsize{0.15--0.20} &\scriptsize{0.20--0.30}
& \scriptsize{0.30--0.40} & \scriptsize{0.40--0.60} & \scriptsize{0.60--0.80} & \scriptsize{0.80--1.20} & \scriptsize{1.20--2.00} & \scriptsize{2.00--4.00} \\ \hline

\scriptsize{0.00--0.01} &1& & & & & & & & & & & & & & \\  
\scriptsize{0.01--0.02} &0.14& 1& & & & & & & & & & & & & \\
\scriptsize{0.02--0.03} &0.16& 0.12& 1& & & & & & & & & & & & \\
\scriptsize{0.03--0.05} &0.20& 0.17& 0.17& 1& & & & & & & & & & & \\
\scriptsize{0.05--0.07} &0.15& 0.12& 0.13& 0.16& 1& & & & & & & & & & \\
\scriptsize{0.07--0.10} &0.19& 0.16& 0.17& 0.21& 0.15& 1& & & & & & & & & \\
\scriptsize{0.10--0.15} &0.14& 0.12& 0.13& 0.16& 0.12& 0.15& 1& & & & & & & & \\
\scriptsize{0.15--0.20} &0.11& 0.10& 0.10& 0.13& 0.10& 0.12& 0.09& 1& & & & & & &\\
\scriptsize{0.20--0.30} &0.15& 0.13& 0.13& 0.17& 0.13& 0.16& 0.13& 0.10& 1& & & & & &\\
\scriptsize{0.30--0.40} &0.10& 0.08& 0.09& 0.11& 0.08& 0.10& 0.08& 0.06& 0.08& 1& & & & &\\
\scriptsize{0.40--0.60} &0.07& 0.06& 0.07& 0.08& 0.06& 0.08& 0.06& 0.05& 0.07& 0.04& 1& & & &\\
\scriptsize{0.60--0.80} &0.05& 0.04& 0.04& 0.05& 0.04& 0.05& 0.04& 0.03& 0.04& 0.03& 0.02& 1& & &\\
\scriptsize{0.80--1.20} &0.05& 0.04& 0.04& 0.05& 0.04& 0.05& 0.04& 0.03& 0.04& 0.03& 0.02& 0.01& 1& &\\
\scriptsize{1.20--2.00} &0.02& 0.02& 0.02& 0.03& 0.02& 0.03& 0.02& 0.02& 0.02& 0.01& 0.01& 0.01& 0.00& 1&\\
\scriptsize{2.00--4.00} &0.005& 0.004& 0.005& 0.006& 0.004& 0.005& 0.004& 0.004& 0.004& 0.002& 0.002& 0.002& 0.002& 0.001& 1\\ \hline

\label{tab:correlPhi}
      \end{tabular}
}
\end{sidewaystable}

\begin{sidewaystable}
 \caption{Correlation coefficients between differential cross-section 
measurements as a function of \PW boson muon $\eta$. The beam energy 
and luminosity 
uncertainties, which are fully correlated between cross-section measurements, are excluded.}
 \resizebox{25cm}{5cm}{
  \begin{tabular}{l l | c c c c c c c c c c c c c c c c | r}

 &$\eta^{\mu}$& \multicolumn{2}{c}{\footnotesize{2--2.25}} & \multicolumn{2}{c}{\footnotesize{2.25--2.5}} & \multicolumn{2}{c}{\footnotesize{2.5--2.75}} & \multicolumn{2}{c}{\footnotesize{2.75--3}} & \multicolumn{2}{c}{\footnotesize{3--3.25}} &\multicolumn{2}{c}{\footnotesize{3.25--3.5}} & \multicolumn{2}{c}{\footnotesize{3.5--4}} & \multicolumn{2}{c}{\footnotesize{4--4.5}}  &  \\ \hline

&\multirow{2}{*}{\rotatebox[origin=c]{0}{\scriptsize{2--2.25}}}&1& & & & & & & & & & & & & & & & $W^{+}$\\
&&0.46& 1& & & & & & & & & & & & & & &  $W^{-}$\\ 
&\multirow{2}{*}{\rotatebox[origin=c]{0}{\scriptsize{2.25--2.5}}}&-0.15& 0.34& 1& & & & & & & & & & & & & & $W^{+}$\\
&&0.30& -0.24& 0.09& 1& & & & & & & & & & & & &  $W^{-}$\\ 
&\multirow{2}{*}{\rotatebox[origin=c]{0}{\scriptsize{2.5--2.75}}}&-0.03& 0.23& 0.27& -0.13& 1& & & & & & & & & & & & $W^{+}$\\
&&0.24& -0.14& -0.21& 0.35& 0.41& 1& & & & & & & & & & & $W^{-}$\\ 
&\multirow{2}{*}{\rotatebox[origin=c]{0}{\scriptsize{2.75--3}}}&0.20& -0.07& -0.10& 0.25& -0.00& 0.21& 1& & & & & & & & & & $W^{+}$\\
&&-0.19& 0.39& 0.46& -0.37& 0.27& -0.24& 0.28& 1& & & & & & & & & $W^{-}$\\ 
&\multirow{2}{*}{\rotatebox[origin=c]{0}{\scriptsize{3--3.25}}}&-0.03& 0.24& 0.28& -0.16& 0.20& -0.08& -0.01& 0.29& 1& & & & & & & & $W^{+}$\\
&&0.31& -0.23& -0.32& 0.46& -0.13& 0.35& 0.25& -0.37& 0.27& 1& & & & & & & $W^{-}$\\ 
&\multirow{2}{*}{\rotatebox[origin=c]{0}{\scriptsize{3.25--3.5}}}&-0.12& 0.29& 0.36& -0.27& 0.22& -0.17& -0.07& 0.40& 0.24& -0.27& 1& & & & & & $W^{+}$\\
&&0.33& -0.32& -0.41& 0.52& -0.19& 0.40& 0.28& -0.47& -0.21& 0.53& -0.07& 1& & & & & $W^{-}$\\ 
&\multirow{2}{*}{\rotatebox[origin=c]{0}{\scriptsize{3.5--4}}}&0.02& 0.15& 0.18& -0.05& 0.15& -0.01& 0.04& 0.17& 0.14& -0.04& 0.15& -0.08& 1& & & & $W^{+}$\\
&&0.21& -0.09& -0.14& 0.30& -0.02& 0.25& 0.19& -0.17& -0.04& 0.30& -0.12& 0.33& 0.45& 1& & & $W^{-}$\\ 
&\multirow{2}{*}{\rotatebox[origin=c]{0}{\scriptsize{4--4.5}}}&-0.01& 0.13& 0.17& -0.08& 0.14& -0.05& -0.01& 0.15& 0.11& -0.07& 0.11& -0.12& 0.09& 0.02& 1& & $W^{+}$\\
&&0.08& 0.01& -0.01& 0.09& 0.03& 0.08& 0.07& -0.02& 0.02& 0.10& -0.02& 0.09& 0.03& 0.10& 0.11& 1& $W^{-}$\\ \hline
&&$W^{+}$ & $W^{-}$ & $W^{+}$ & $W^{-}$ & $W^{+}$ & $W^{-}$ & $W^{+}$ & $W^{-}$ & $W^{+}$ & $W^{-}$ & $W^{+}$ & $W^{-}$ & $W^{+}$ & $W^{-}$ & $W^{+}$ & $W^{-}$ & \\

\label{tab:correlEta}
      \end{tabular}
}
\end{sidewaystable}

\begin{sidewaystable}
 \caption{Correlation coefficients between differential cross-section 
measurements as a function of $y_{Z}$ and \PW boson muon $\eta$. The \lhc 
beam energy and luminosity 
uncertainties, which are fully correlated between cross-section measurements, are excluded.}
 \resizebox{25cm}{4cm}{
  \begin{tabular}{l l | c c c c c c c c c c c c c c c c c c | r}

 && \multicolumn{18}{c}{$y_{\PZ}$} & \\
 && \scriptsize{2--2.125} & \scriptsize{2.125--2.25} & \scriptsize{2.25--2.375} & \scriptsize{2.375--2.5} & \scriptsize{2.5--2.625} & \scriptsize{2.625--2.75} & 
\scriptsize{2.75--2.875} & \scriptsize{2.875--3} & \scriptsize{3--3.125} &
\scriptsize{3.125--3.25} & \scriptsize{3.25--3.375} & \scriptsize{3.375--3.5} & \scriptsize{3.5--3.625} &
\scriptsize{3.625--3.75} &
\scriptsize{3.75--3.875} & \scriptsize{3.875--4} & \scriptsize{4--4.25}  & \scriptsize{4.25--4.5} &  \\ \hline

\multirow{16}{*}{\rotatebox[origin=c]{0}{$\eta^{\mu}$}}&\multirow{2}{*}{\rotatebox[origin=c]{0}{\scriptsize{2--2.25}}}&0.24& 0.25& 0.19& 0.19& 0.17& 0.16& 0.15& 0.13& 0.12& 0.10& 0.08& 0.06& 0.05& 0.04& 0.03& 0.02& 0.01& $-$& $W^{+}$\\
&&0.22& 0.23& 0.17& 0.17& 0.16& 0.15& 0.13& 0.12& 0.11& 0.09& 0.07& 0.05& 0.04& 0.04& 0.03& 0.02& 0.01& $-$& $W^{-}$\\ 
&\multirow{2}{*}{\rotatebox[origin=c]{0}{\scriptsize{2.25--2.5}}}&0.03& 0.11& 0.13& 0.14& 0.12& 0.11& 0.10& 0.10& 0.09& 0.07& 0.06& 0.04& 0.03& 0.02& 0.01& 0.01& 0.01& $-$& $W^{+}$\\
&&0.03& 0.10& 0.12& 0.13& 0.11& 0.10& 0.10& 0.09& 0.08& 0.07& 0.06& 0.04& 0.03& 0.02& 0.01& 0.01& 0.00& $-$& $W^{-}$\\ 
&\multirow{2}{*}{\rotatebox[origin=c]{0}{\scriptsize{2.5--2.75}}}&0.03& 0.04& 0.06& 0.10& 0.10& 0.10& 0.09& 0.09& 0.08& 0.07& 0.07& 0.05& 0.04& 0.03& 0.02& 0.01& 0.01& $-$& $W^{+}$\\
&&0.03& 0.03& 0.06& 0.10& 0.09& 0.09& 0.08& 0.08& 0.08& 0.07& 0.06& 0.05& 0.04& 0.02& 0.02& 0.01& 0.01& $-$& $W^{-}$\\ 
&\multirow{2}{*}{\rotatebox[origin=c]{0}{\scriptsize{2.75--3}}}&0.03& 0.04& 0.04& 0.07& 0.09& 0.10& 0.09& 0.09& 0.09& 0.08& 0.07& 0.06& 0.05& 0.04& 0.02& 0.01& 0.01& $-$& $W^{+}$\\
&&0.02& 0.03& 0.03& 0.05& 0.07& 0.07& 0.07& 0.07& 0.06& 0.06& 0.05& 0.04& 0.04& 0.03& 0.01& 0.01& 0.01& $-$& $W^{-}$\\ 
&\multirow{2}{*}{\rotatebox[origin=c]{0}{\scriptsize{3--3.25}}}&0.04& 0.04& 0.04& 0.06& 0.08& 0.10& 0.10& 0.10& 0.09& 0.09& 0.08& 0.07& 0.06& 0.05& 0.03& 0.01& 0.01& $-$& $W^{+}$\\
&&0.03& 0.04& 0.03& 0.05& 0.06& 0.08& 0.08& 0.08& 0.08& 0.07& 0.07& 0.06& 0.05& 0.04& 0.02& 0.01& 0.01& $-$& $W^{-}$\\ 
&\multirow{2}{*}{\rotatebox[origin=c]{0}{\scriptsize{3.25--3.5}}}&0.02& 0.03& 0.03& 0.04& 0.04& 0.05& 0.06& 0.06& 0.06& 0.06& 0.05& 0.05& 0.04& 0.03& 0.02& 0.01& 0.01& $-$& $W^{+}$\\
&&0.02& 0.02& 0.02& 0.03& 0.03& 0.04& 0.05& 0.05& 0.05& 0.04& 0.04& 0.03& 0.03& 0.02& 0.02& 0.01& 0.00& $-$& $W^{-}$\\ 
&\multirow{2}{*}{\rotatebox[origin=c]{0}{\scriptsize{3.5--4}}}&0.03& 0.03& 0.03& 0.04& 0.04& 0.05& 0.06& 0.08& 0.09& 0.08& 0.08& 0.07& 0.07& 0.05& 0.04& 0.03& 0.02& $-$& $W^{+}$\\
&&0.03& 0.03& 0.03& 0.04& 0.04& 0.05& 0.06& 0.08& 0.09& 0.09& 0.08& 0.07& 0.07& 0.05& 0.04& 0.03& 0.02& $-$& $W^{-}$\\ 
&\multirow{2}{*}{\rotatebox[origin=c]{0}{\scriptsize{4--4.5}}}&0.02& 0.02& 0.02& 0.02& 0.02& 0.02& 0.02& 0.02& 0.03& 0.04& 0.05& 0.05& 0.06& 0.06& 0.04& 0.03& 0.03& $-$& $W^{+}$\\
&&0.02& 0.02& 0.02& 0.03& 0.03& 0.03& 0.03& 0.03& 0.04& 0.05& 0.07& 0.07& 0.07& 0.07& 0.05& 0.04& 0.03& $-$& $W^{-}$\\ \hline

\hline

\label{tab:correlWZ}
      \end{tabular}
}
\end{sidewaystable}

\clearpage


\addcontentsline{toc}{section}{References}
\setboolean{inbibliography}{true}
\bibliographystyle{LHCb}
\bibliography{main,LHCb-PAPER,LHCb-CONF,LHCb-DP,LHCb-TDR}

\newpage


\centerline{\large\bf LHCb collaboration}
\begin{flushleft}
\small
R.~Aaij$^{38}$, 
B.~Adeva$^{37}$, 
M.~Adinolfi$^{46}$, 
A.~Affolder$^{52}$, 
Z.~Ajaltouni$^{5}$, 
S.~Akar$^{6}$, 
J.~Albrecht$^{9}$, 
F.~Alessio$^{38}$, 
M.~Alexander$^{51}$, 
S.~Ali$^{41}$
% ...

{\footnotesize \it
% ...
$ ^{5}$Clermont Universit\'{e}, Universit\'{e} Blaise Pascal, CNRS/IN2P3, LPC, Clermont-Ferrand, France\\
$ ^{6}$CPPM, Aix-Marseille Universit\'{e}, CNRS/IN2P3, Marseille, France\\
% ...
$ ^{9}$Fakult\"{a}t Physik, Technische Universit\"{a}t Dortmund, Dortmund, Germany\\
% ...
$ ^{37}$Universidad de Santiago de Compostela, Santiago de Compostela, Spain\\
$ ^{38}$European Organization for Nuclear Research (CERN), Geneva, Switzerland\\
% ...
$ ^{41}$Nikhef National Institute for Subatomic Physics, Amsterdam, The Netherlands\\
% ...
$ ^{46}$H.H. Wills Physics Laboratory, University of Bristol, Bristol, United Kingdom\\
% ...
$ ^{51}$School of Physics and Astronomy, University of Glasgow, Glasgow, United Kingdom\\
$ ^{52}$Oliver Lodge Laboratory, University of Liverpool, Liverpool, United Kingdom\\
% ...
}
\end{flushleft}

\newpage

\end{document}